# Transformation-Based Bottom-Up Computation of the Well-Founded Model

STEFAN BRASS

*School of Information Sciences, University of Pittsburgh,*
*135 North Bellefield Ave., Pittsburgh, PA 15260, USA,*
*e-mail: sbrass@sis.pitt.edu*

JÜRGEN DIX

*The University of Manchester, Dept. of Computer Science,*
*Oxford Road, Manchester, M13 9PL, UK,*
*email: jdix@cs.man.ac.uk*

BURKHARD FREITAG and ULRICH ZUKOWSKI

*Universität Passau, Fakultät für Mathematik und Informatik,*
*D-94030 Passau, Germany,*
*email: freitag,zukowski@fmi.uni-passau.de*

## Abstract

We present a framework for expressing bottom-up algorithms to compute the well-founded model of non-disjunctive logic programs. Our method is based on the notion of conditional facts and elementary program transformations studied by BRASS and DIX in (Brass & Dix, 1994; Brass & Dix, 1999) for disjunctive programs. However, even if we restrict their framework to nondisjunctive programs, their "residual program" can grow to exponential size, whereas for function-free programs our "program remainder" is always polynomial in the size of the extensional database (EDB).

We show that particular orderings of our transformations (we call them *strategies*) correspond to well-known computational methods like the alternating fixpoint approach (Van Gelder, 1989; Van Gelder, 1993), the well-founded magic sets method (Kemp *et al.*, 1995) and the magic alternating fixpoint procedure (Morishita, 1996). However, due to the confluence of our calculi (first noted in (Brass & Dix, 1998)), we come up with computations of the well-founded model that are provably better than these methods.

In contrast to other approaches, our transformation method treats magic set transformed programs correctly, i.e. it always computes a relevant part of the well-founded model of the original program.

These results show that our approach is a valuable tool to analyze, compare, and optimize existing evaluation methods or to create new strategies that are automatically proven to be correct if they can be described by a sequence of transformations in our framework. We have also developed a prototypical implementation. Experiments illustrate that the theoretical results carry over to the implemented prototype and may be used to optimize real life systems.

## 1 Introduction

The next generation of deductive database systems (see Ramakrishnan's article in (Apt *et al.*, 1999)) will probably support the full class of normal programs. It seems also very likely that the well-founded semantics will be chosen by many system designers, because it has a unique model. Furthermore, the time complexity of computing the well-founded model of a given intensional database (IDB) is polynomial in the size of the extensional database (EDB) (Van Gelder *et al.*, 1991). In contrast, it has been shown by Marek and Truszczynski (Marek & Truszczynski, 1991) that even for a propositional logic program $P$, determining whether $P$ has a stable model is NP-complete. Extensions of deductive database systems that can deal with the well-founded semantics are already realized in XSB (Chen *et al.*, 1995; Chen & Warren, 1996; Sagonas *et al.*, 1994) and announced for LOLA (Freitag *et al.*, 1991; Zukowski & Freitag, 1996a; Zukowski & Freitag, 1996b; Zukowski & Freitag, 1997).

The SLG-resolution of Chen and Warren (Chen *et al.*, 1995; Chen & Warren, 1996), as implemented in the XSB system, is the most prominent *top-down* method for the computation of the well-founded model of a normal program. In this paper we characterize "good" *bottom-up* methods in terms of elementary program transformations. Essentially, the bottom-up algorithms that compute the well-founded model of a normal program are:

- The alternating fixpoint approach, introduced by Van Gelder (Van Gelder, 1989; Van Gelder, 1993) and further developed by Kemp, Stuckey and Srivastava (Kemp *et al.*, 1995). Many other bottom-up methods (Kemp *et al.*, 1995; Morishita, 1996; Stuckey & Sudarshan, 1997) are based on this approach.
- The computation of the residual program, as suggested by Bry (Bry, 1989; Bry, 1990) and independently by Dung/Kanchanasut (Dung & Kanchansut, 1989a; Dung & Kanchansut, 1989b), and extended by Brass and Dix (Brass & Dix, 1999; Brass & Dix, 1998) for disjunctive logic programs.

The alternating fixpoint procedure is known to have efficiency problems in the sense that it needs quadratic evaluation time for programs that could be easily computed in linear time w.r.t. their size. In every iteration many facts have to be recomputed. The residual program approach avoids recomputations of this kind, and thus needs only linear time for many programs the alternating fixpoint approach needs quadratic time for. But it is possible that the residual program can grow to exponential size while for function-free programs the alternating fixpoint approach guarantees a number of derived facts that is polynomial in the size, i.e. the number of tuples, of the extensional database (EDB). However, the residual program contains important information which is not provided by the alternating fixpoint method.

It is natural to ask for an algorithm which combines the advantages of both bottom-up approaches. In this paper, we present a framework for the computation of the well-founded model that is based on the residual program method but guarantees polynomial complexity like the alternating fixpoint procedure. The residual program method is based on elementary program transformations and the concept of *conditional facts*. Conditional facts are ground rules having those negative literals in their bodies that can not (yet) be resolved because their complement is not (yet) known to be true or false. This can be regarded as a non-procedural equivalent of the *delay* operation needed in SLG-resolution (Chen *et al.*, 1995; Chen & Warren, 1996). Like in SLG-Resolution, the key idea to avoid the exponential blow-up of the residual program is to delay not only negative literals but also positive literals which depend on delayed negative literals. The transformation-based *program remainder* method which we propose in this paper can be seen as a bottom-up equivalent of SLG-resolution for the case of range-restricted function-free programs.

We also introduce the important notion of a *regular strategy expression*, which is used to define different evaluation methods in a simple and declarative way. Various such strategies are shown to correspond exactly to the alternating fixpoint computation ((Van Gelder, 1989; Van Gelder, 1993)), the well-founded magic sets method (Kemp *et al.*, 1995) and the magic alternating fixpoint procedure (Morishita, 1996).

But our method is not applicable only to range-restricted programs. In many cases, the magic set transformed version of a non-range-restricted program w.r.t. a given query is range-restricted. Although it is known that the magic set transformation (Beeri & Ramakrishnan, 1991) has problems in the context of the well-founded semantics our method treats magic set transformed programs correctly. Moreover, we show that our approach is guaranteed to need not more work than the well-founded magic sets method (Kemp *et al.*, 1995) or the magic alternating fixpoint (Morishita, 1996) and is much more efficient for many programs.

The rest of the paper is organized as follows. After introducing preliminaries in Section 2, we recall the alternating fixpoint procedure in Section 3. In Section 4, we introduce our program transformation approach for ground programs. In Section 5 we present a method that performs an intelligent grounding. Section 6 compares our approach to the alternating fixpoint procedure. Section 7 introduces a regular expression like syntax to express and compare different evaluation strategies. In Section 8 we extend our transformation approach to handle magic set transformed programs. Section 9 discusses our approach and compares it with related work. Section 10 concludes the paper.

## 2 Preliminaries

A rule is of the form $A \leftarrow L_1 \wedge \cdots \wedge L_n$, where the head $A$ is an atom and each body literal $L_i$ is a positive literal, i.e. an atom, or a negative literal $L_i = \textbf{not}\, B$. We treat the rule body as a set of literals and write also $A \leftarrow \mathcal{B}$ with $\mathcal{B} = \{L_1, \ldots, L_n\}$. Consequently, a fact $A$ is represented by the rule $A \leftarrow \emptyset$. We also use the notation $A \leftarrow \mathcal{B} \wedge \textbf{not}\, \mathcal{C}$ with atom sets $\mathcal{B}$ and $\mathcal{C}$.

A program is a set of rules as introduced above. We consider normal, i.e., non-disjunctive, logic programs without function symbols. We assume that all rules are range-restricted, i.e. that each variable of the rule appears also in a positive body literal. The *size* of a program is the number of literal occurrences.

Let $P$ be a program. $BASE(P)$ denotes the Herbrand base of $P$, i.e. the set of all ground atoms which can be built using the predicates and constants occurring in $P$. It might happen that our transformations eliminate some constants and predicates from the program, but we use the Herbrand base $BASE(P)$ of the original program. We write $ground(P)$ for the Herbrand instantiation of a program $P$. For a ground program $P$ we define the following sets:

$$
\begin{aligned}
facts(P) &:= \{A \in BASE(P) \mid (A \leftarrow \emptyset) \in P\}, \\
heads(P) &:= \{A \in BASE(P) \mid \text{there is a } \mathcal{B} \text{ such that } (A \leftarrow \mathcal{B}) \in P\}.
\end{aligned}
$$

The complement of a literal $L$ is denoted by $\sim L$, i.e., $\sim(B) = \mathbf{not}\,B$ and, conversely, $\sim(\mathbf{not}\,B) = B$. For a set $S$ of literals, $\sim S$ denotes the set of the complements of the literals in $S$. $S$ is *consistent* if and only if $S \cap \sim S = \emptyset$. For a set $S$ of literals we define the following sets:

$$
\begin{aligned}
pos(S) &:= \{A \in S \mid A \text{ is a positive literal }\}, \\
neg(S) &:= \{A \mid \mathbf{not}\,A \in S\}.
\end{aligned}
$$

Let $P$ be a logic program. A *partial interpretation* for $P$ is a consistent set $I$ of ground literals such that

$$pos(I) \cup neg(I) \subseteq BASE(P),$$

i.e., its set of atoms is a subset of the Herbrand base of $P$. A *total interpretation* for $P$ is a partial interpretation $I$ such that

$$A \in BASE(P) \implies A \in I \text{ or } \mathbf{not}\,A \in I,$$

i.e., for each atom A of the Herbrand base of $P$ either $A$ or its complement is contained in $I$. Two partial interpretations are said to agree on an atom $Q$ if they contain the same ground instances of $Q$ and $\mathbf{not}\,Q$.

For an operator $T$ we define $T \uparrow 0 := \emptyset$ and $T \uparrow i := T(T \uparrow i-1)$, for $i > 0$. $\mathrm{lfp}(T)$ denotes the least fixpoint of $T$, i.e. the smallest set $S$ such that $T(S) = S$. We rely on the definition of the *well-founded partial model* $W_P^*$ of $P$ as given in (Van Gelder *et al.*, 1991). An operational characterization of $W_P^*$ will be given in Section 3. Further, we assume familiarity with the concepts of the *magic set transformation* (Beeri & Ramakrishnan, 1991).

## 3 The Alternating Fixpoint Procedure

Let us recall the definition of the alternating fixpoint procedure. We introduce an extended version of the immediate consequence operator that uses two different sets of facts for positive and negative subgoals, respectively. Actually, this is an adaptation of the stability transformation found in (Van Gelder, 1989; Van Gelder, 1993) to our purposes. This operator has been introduced and investigated by PRZY-MUSINSKI in (Przymusinski, 1989b; Przymusinski, 1990; Przymusinski, 1991).

*Definition 1 (Extended Immediate Consequence Operator)*

Let $P$ be a normal logic program. Let $I$ and $J$ be sets of ground atoms. The set

$T_{P,J}(I)$ of *immediate consequences of $I$ w.r.t. $P$ and $J$* is defined as follows:

$$T_{P,J}(I) := \{A \mid \text{there is } A \leftarrow \mathcal{B} \in ground(P) \text{ with } \begin{array}{l} pos(\mathcal{B}) \subseteq I \text{ and} \\ neg(\mathcal{B}) \cap J = \emptyset\}. \end{array}$$

If $P$ is definite, the set $J$ is not needed and we obtain the standard immediate consequence operator $T_P$ by $T_P(I) = T_{P,\emptyset}(I)$.

$T_{P,J}$ checks negative subgoals against the set of possibly true atoms that is supplied as the argument $J$. This allows the following elegant formulation of the alternating fixpoint procedure.

*Definition 2 (Alternating Fixpoint Procedure)*
Let $P$ be a normal logic program. Let $P^+$ denote the subprogram consisting of the definite rules of $P$. Then the sequence $(K_i, U_i)_{i \geq 0}$ with sets $K_i$ of true (known) facts and $U_i$ of possible (unknown) facts is defined by:

$$\begin{array}{llllll} K_0 & := & \text{lfp}(T_{P^+}) & U_0 & := & \text{lfp}(T_{P,K_0}) \\ i > 0: & K_i & := & \text{lfp}(T_{P,U_{i-1}}), & U_i & := & \text{lfp}(T_{P,K_i}). \end{array}$$

The computation terminates when the sequence becomes stationary, i.e., when a fixpoint is reached in the sense that

$$(K_j, U_j) = (K_{j+1}, U_{j+1}).$$

This computation schema is called the *Alternating Fixpoint Procedure* (AFP).

*Theorem 3 (Correctness of AFP (Van Gelder, 1993))*
Let the sequence $(K_i, U_i)_{i \geq 0}$ be defined as above. Then there is a $j \geq 0$ such that $(K_j, U_j) = (K_{j+1}, U_{j+1})$. The well-founded model $W_P^*$ of $P$ can be directly derived from the fixpoint $(K_j, U_j)$, i.e.,

$W_P^* = \{L \mid L$ is a positive ground literal and $L \in K_j$ or
$\qquad\qquad L$ is a negative ground literal $\textbf{not}\, A$ and $A \in BASE(P) - U_j\}$.

The following results which follow easily from the anti-monotonicity of $T_{P,J}$ w.r.t. $J$ are also stated in (Kemp *et al.*, 1995).

*Lemma 4 (Monotonicity)*
Let the sequence $(K_i, U_i)_{i \geq 0}$ be defined as above. Then the following holds for $i \geq 0$:
$K_i \subseteq K_{i+1}, \quad U_i \supseteq U_{i+1}, \quad K_i \subseteq U_i$.

Whereas the sets $K_i$ of true facts can be computed incrementally, the sets $U_i$ of possible facts have to be recomputed in every iteration step. It is well-known that this leads to many unnecessary recomputations even for simple programs.

*Example 5 (Quadratic Time Complexity of AFP)*
Consider the following logic program taken from (Kemp *et al.*, 1995; Morishita, 1996)

$$\begin{array}{lll} p(X) & \leftarrow & t(X, Y, Z), \textbf{not}\, p(Y), \textbf{not}\, p(Z). \\ p(X) & \leftarrow & p_0(X). \end{array}$$

and the following base facts:

$$p_0(c_2), t(a, a, b_1), t(b_1, c_1, b_2), t(b_2, c_2, b_3), \ldots, t(b_n, c_n, b_{n+1}).$$

Due to the linear character of the graph of dependencies between the atoms $p(b_1)$ to $p(b_n)$ we would expect that it is possible to compute the well-founded model of this program in linear time w.r.t. $n$. However, the alternating fixpoint procedure needs $\frac{n}{2}$ iterations, each deriving a number of facts that is linear in $n$. So the total execution time is quadratic in $n$. Note that this effect becomes evident also for simpler programs, but we have chosen this program because it is interesting in the context of the magic set transformation and we can use the same example throughout the whole paper.

## 4 Basic Ground Transformations

Brass and Dix (Brass & Dix, 1994; Brass & Dix, 1999; Brass & Dix, 1997) have introduced a framework for studying and computing negation semantics by means of elementary program transformations. Based on this framework we introduce our transformation approach.

*Definition 6 (Program Transformation)*
A *program transformation* is a relation $\mapsto$ between ground logic programs.

The following two transformations reduce negative body literals if their truth value is obvious. These transformation are also used in (Brass & Dix, 1999).

*Definition 7 (Positive Reduction)*
Let $P_1$ and $P_2$ be ground programs. $P_2$ results from $P_1$ by *positive reduction* $(P_1 \mapsto_P P_2)$ iff there is a rule $A \leftarrow \mathcal{B}$ in $P_1$ and a negative literal $\mathbf{not}\, B \in \mathcal{B}$ such that there is no rule about $B$ in $P_1$, i.e., $B \notin heads(P_1)$, and $P_2 = (P_1 - \{A \leftarrow \mathcal{B}\}) \cup \{A \leftarrow (\mathcal{B} - \{\mathbf{not}\, B\})\}$.

*Definition 8 (Negative Reduction)*
Let $P_1$ and $P_2$ be ground programs. $P_2$ results from $P_1$ by *negative reduction* $(P_1 \mapsto_N P_2)$ iff there is a rule $A \leftarrow \mathcal{B}$ in $P_1$ and a negative literal $\mathbf{not}\, B \in \mathcal{B}$ such that $B$ appears as a fact in $P_1$, i.e., $B \in facts(P_1)$, and $P_2 = P_1 - \{A \leftarrow \mathcal{B}\}$.

In (Brass & Dix, 1999) positive body literals are resolved by unfolding. However, it is well-known that by unrestricted unfolding the resulting program can grow to a size that is exponential in the size of the original program (Chen & Warren, 1996; Zukowski *et al.*, 1997).

*Example 9 (Exponential Residual Program)*
Consider the following program.

$$
\begin{aligned}
p(0). &\\
p(X) &\leftarrow p(Y), succ(Y, X), \mathbf{not}\, q(Y).\\
p(X) &\leftarrow p(Y), succ(Y, X), \mathbf{not}\, r(Y).\\
q(X) &\leftarrow succ(X, \_), \mathbf{not}\, q(X).\\
r(X) &\leftarrow succ(X, \_), \mathbf{not}\, r(X).
\end{aligned}
$$

$$succ(0,1).\ \ldots\ succ(n-1,n).$$

The following diagram depicts all possible paths on which a $p$-fact can be derived:

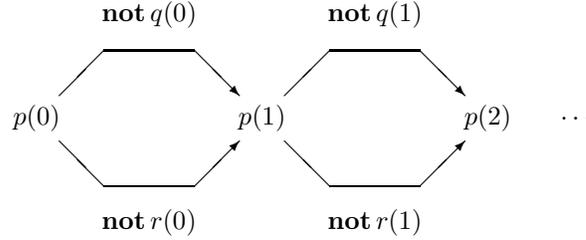

By allowing unrestricted unfolding of all positive body literals we get the following program that is called *residual program*:

$$p(0).$$

$$
\begin{aligned}
p(1) &\leftarrow \mathbf{not}\,q(0).\\
p(1) &\leftarrow \mathbf{not}\,r(0).\\[4pt]
p(2) &\leftarrow \mathbf{not}\,q(0), \mathbf{not}\,q(1).\\
p(2) &\leftarrow \mathbf{not}\,q(0), \mathbf{not}\,r(1).\\
p(2) &\leftarrow \mathbf{not}\,r(0), \mathbf{not}\,q(1).\\
p(2) &\leftarrow \mathbf{not}\,r(0), \mathbf{not}\,r(1).
\end{aligned}
$$

$$\vdots$$

It is obvious that every possible path is encoded in the residual program. Each possible fact $p(n)$ is represented by $2^n$ conditional facts.

To achieve a polynomial complexity we introduce the following two transformations that are special cases of unfolding and reduce positive body literals only if their truth value is obvious.

**Definition 10 (*Success*)**
Let $P_1$ and $P_2$ be ground programs. $P_2$ results from $P_1$ by *success* $(P_1 \mapsto_S P_2)$ iff there is a rule $A \leftarrow \mathcal{B}$ in $P_1$ and a positive literal $B \in \mathcal{B}$ such that $B \in facts(P_1)$, and $P_2 = \big(P_1 - \{A \leftarrow \mathcal{B}\}\big) \cup \big\{A \leftarrow (\mathcal{B} - \{B\})\big\}$.

**Definition 11 (*Failure*)**
Let $P_1$ and $P_2$ be ground programs. $P_2$ results from $P_1$ by *failure* $(P_1 \mapsto_F P_2)$ iff there is a rule $A \leftarrow \mathcal{B}$ in $P_1$ and a positive literal $B \in \mathcal{B}$ such that there is no rule about $B$ in $P_1$, i.e., $B \notin heads(P_1)$, and $P_2 = P_1 - \{A \leftarrow \mathcal{B}\}$.

**Definition 12**
Let $\mapsto_{PSNF}$ be the rewriting system consisting of the above four transformations, i.e. $\mapsto_{PSNF} := \mapsto_P \cup \mapsto_N \cup \mapsto_S \cup \mapsto_F$.

This rewriting system is to be applied to a program as long as it is applicable, i.e., until a normal form w.r.t. these transformations is reached.

*Definition 13 (Normal Form)*
A program $P'$ is a *normal form* of a program $P$ w.r.t. a transformation $\mapsto$ iff $P \mapsto^* P'$, and $P'$ is irreducible, i.e. there is no program $P''$ with $P' \mapsto P''$.

When a normal form $P'$ is reached, the result of the transformation is interpreted in the following way: atoms appearing as facts in $P'$ are true in $P'$, atoms from the Herbrand base of $P$ not appearing in any rule head of $P'$ are false in $P'$, all remaining atoms are undefined in $P'$.

*Definition 14 (Known Literals)*
Let $P$ be a ground program and let $S$ be a set of ground atoms. The set of positive and negative ground literals with atoms in $S$ having an obvious truth value in $P$ is denoted by $known_S(P)$:

$$
\begin{aligned}
known_S(P) := \{ L \mid\ &L \in S \text{ is a positive ground literal} \\
&\text{and } L \in S \cap facts(P) \text{ or} \\
&L \text{ is a negative ground literal } \textbf{not } A \\
&\text{and } A \in S - heads(P) \}.
\end{aligned}
$$

The next theorem states that the four transformations *success*, *failure*, *positive* and *negative reduction* correspond exactly to the Fitting operator $\Phi_P$, which is the procedural counterpart of the 3-valued completion $comp_3$ (we refer to (Fitting, 1985; Dix, 1995a; Dix, 1995b) for a detailed exposition of $\Phi_P$).

*Theorem 15 ($\mapsto_{PSNF}$ corresponds to $\mathrm{lfp}(\Phi_P)$)*
The rewriting system $\mapsto_{PSNF}$ is confluent and strongly terminating. With strongly terminating, we mean that not only there exists a terminating sequence for a given program, but any rewriting sequence is terminating. It therefore induces a normal form $fitt(P)$ for any program $P$. The known literals with respect to this normal form constitute exactly the least fixpoint $\mathrm{lfp}(\Phi_P)$ of Fitting's operator $\Phi_P$:

$$
\mathrm{lfp}(\Phi_P) = known_{BASE(P)}\big(fitt(P)\big).
$$

*Proof*
See Appendix A.

Note that for a given ground program $P$ the normal form $fitt(P)$ can be computed in a time linear in the size of $P$. An efficient algorithm based on the results of (Dowling & Gallier, 1984) can be found in (Niemelä & Simons, 1996). The four parts of this implementation, that reduce body literals, correspond directly to the four basic transformations defined above. Thus, that algorithm can be seen as an efficient implementation of our rewriting system $\mapsto_{PSNF}$. For many programs, e.g. for programs without positive loops, the Fitting operator computes the well-founded model. This positive effect of the Fitting operator has also been pointed out in (Berman *et al.*, 1995; Subrahmanian *et al.*, 1995). However, positive loops have to be detected and removed separately.

*Example 16* (*Positive Loop*)

Consider the following program *Loop*:

$$
\begin{aligned}
p.&\\
q &\leftarrow \mathbf{not}\, p.\\
q &\leftarrow r.\\
r &\leftarrow q.
\end{aligned}
$$

We can apply *negative reduction* to delete $q \leftarrow \mathbf{not}\, p$, since $\mathbf{not}\, p$ is obviously false. But $\mapsto_{PSNF}$ does not allow to delete $q \leftarrow r$ and $r \leftarrow q$.

To remove positive loops we introduce the following transformation.

*Definition 17* (*Loop Detection*)

Let $P_1$ and $P_2$ be ground programs. $P_2$ results from $P_1$ by *loop detection* ($P_1 \mapsto_L P_2$) iff there is a set $\mathcal{A}$ of ground atoms such that

1. for each rule $A \leftarrow \mathcal{B}$ in $P_1$, if $A \in \mathcal{A}$, then $\mathcal{B} \cap \mathcal{A} \neq \emptyset$,
2. $P_2 := \{A \leftarrow \mathcal{B} \in P_1 \mid A \notin \mathcal{A}\}$,
3. $P_1 \neq P_2$.

In our earlier papers, we have defined *loop detection* such that all rules are deleted with a positive body atom in $\mathcal{A}$. This has the disadvantage, however, that *loop detection* is not local to SCCs. For instance, in $P = \{p \leftarrow q,\ q \leftarrow p\} \cup \{r \leftarrow q\}$, the old version of *loop detection* would delete all three rules. But then the useful Lemma 18 would not hold.

In contrast to the four transformations of $\mapsto_{PSNF}$ that can be applied locally, for *loop detection* we need a more global perspective on the program. The set $\mathcal{A}$ is the *greatest unfounded set* of $P_1$ (w.r.t. the empty interpretation $\emptyset$, cf. (Van Gelder *et al.*, 1991)). The greatest unfounded set consists of all positive ground atoms which are not possibly true, i.e., cannot be derived by assuming all negative literals to be true. According to Definition 1, the extended immediate consequence operator $T_{P_1, \emptyset}$ computes the possibly true atoms. Therefore, the greatest unfounded set is given by $BASE(P) - \mathrm{lfp}(T_{P_1, \emptyset})$. This set has the property required for $\mathcal{A}$ in Definition 17:

*Lemma 18* (*Loop Detection*)

Let $P_1$ be a ground program and let

$$
P_2 = \big\{ (A \leftarrow \mathcal{B}) \in P_1 \;\big|\; A \in \mathrm{lfp}(T_{P_1, \emptyset}) \big\}.
$$

Then

1. $P_1 \mapsto_L P_2$ (unless $P_2 = P_1$), and
2. $P_2$ is irreducible w.r.t. $\mapsto_L$.

*Proof*

See Appendix A.

As shown in (Niemelä & Simons, 1996), the set $\text{lfp}(T_{P_1,\emptyset})$ of possible atoms can be computed by the linear time algorithm of (Dowling & Gallier, 1984). As a consequence, also the operation of *loop detection* can be performed in time that is linear in the size of $P_1$. We compute $\text{lfp}(T_{P_1,\emptyset})$ and delete all rules $A \leftarrow \mathcal{B}$ with $A \notin \text{lfp}(T_{P_1,\emptyset})$.

*Example 19* (*Loop Detection*)
Consider again the program *Loop* of Example 16. After deleting the second rule by *negative reduction, loop detection* can be applied using the unfounded set $\mathcal{A} = \{q, r\}$ to delete the third and fourth rule.

Since more than one application of loop detection may be needed for some programs, in our final rewriting system we have to iterate all five transformations.

*Definition 20*
Let $\mapsto_X$ denote our final rewriting system:

$$\mapsto_X := \mapsto_P \cup \mapsto_N \cup \mapsto_S \cup \mapsto_F \cup \mapsto_L.$$

Thus $\mapsto_X$ can be seen as the extension of Fittings least fixpoint operator $\Phi_P$ by *loop detection* (see Theorem 15). A normal form w.r.t. the rewriting system $\mapsto_X$ is called a program remainder:

*Definition 21* (*Program Remainder*)
Let $P$ be a program. Let the program $\widehat{P}$ be a normal form of $P$ w.r.t. $\mapsto_X$. Then $\widehat{P}$ is called a *program remainder* of $P$.

*Lemma 22* (*Termination*)
Every program $P$ has a program remainder, i.e. the rewriting system $\mapsto_X$ is terminating.

*Proof*
All our transformations strictly reduce the total number of occurring literals. So if $P_0 := ground(P)$ and $P_i$ is any program with $P_{i-1} \mapsto_X P_i$, $i \geq 1$, then we must reach a program $P_n$ in which no further transformation can be applied. This is a program remainder of $P$. $\square$

*Theorem 23* (*Computation of Well-Founded Semantics*)
For every program $P$ and remainder $\widehat{P}$, the well-founded model $W_P^*$ of $P$ satisfies exactly those positive and negative literals which are immediately obvious in $\widehat{P}$, i.e.

$$W_P^* = known_{BASE(P)}(\widehat{P}).$$

*Proof*
See Appendix A.

From this it is also simple to derive an explicit characterization of the program remainder: From the Herbrand instantiation of the program we obtain the program remainder by

1. deleting every rule instance with a body literal which is false in the well-founded model, and

2. removing from the remaining rule instances the body literals which are true in the well-founded model.

This is a kind of SMALL CAPS GELFOND/LIFSCHITZ-transformation (Gelfond & Lifschitz, 1988).

*Theorem 24 (Unique Program Remainder)*

Let $\widehat{P}$ be a remainder of a program $P$. The following holds:

$$\widehat{P} = \big\{ A \leftarrow (\mathcal{B} - W_P^*) \mid A \leftarrow \mathcal{B} \in \ \mathit{ground}(P)$$
$$\text{and for every } B \in \mathcal{B}: \ \sim\! B \notin W_P^* \big\}.$$

Thus the remainder of a program is uniquely determined. In the following we will denote the program remainder of $P$ by $\widehat{P}$.

*Proof*

See Appendix A.

*Corollory 25 (Confluence)*

The rewriting system $\mapsto_X$ is confluent, strongly terminating and the program remainder $\widehat{P}$ is the unique normal form of $\mathit{ground}(P)$ w.r.t. $\mapsto_X$.

Remember that, as a consequence of unlimited unfolding, the residual program, i.e. the normal form of the rewriting system of (Brass & Dix, 1999), can grow to a size that is exponential in the size of the original program. In our approach the program does not grow at all since all transformations strictly reduce the size of the program.

*Lemma 26 (Quadratic Time complexity)*

Let $P$ be a ground program. Let $n$ be the size of $P$, i.e., the number of occurrences of literals in $P$. Then $P$ can be transformed into its remainder $\widehat{P}$ in time $O(n^2)$.

*Proof*

As shown in (Niemelä & Simons, 1996) the normal form of $P$ w.r.t. $\mapsto_{PSNF}$ can be computed in time $O(n)$. Also one application of *loop detection* can be performed in time $O(n)$. Since by each application of *loop detection* the size of the program is strictly reduced, *loop detection* can be applied at most $n$ times. Altogether, the remainder $\widehat{P}$ of $P$ is reached in time $O(n^2)$. $\quad\square$

Note that quadratic time is needed in the worst case if *loop detection* is to be applied $n$ times. Time complexity can be improved up to linear time, if the number of applications of *loop detection* can be reduced, depending on the dependencies in $P$.

*Example 27 (Linear Time Complexity)*

Consider the following ground program

$$
\begin{aligned}
p(a) &\leftarrow \mathbf{not}\, p(a), \mathbf{not}\, p(b_1). \\
p(b_1) &\leftarrow \mathbf{not}\, p(c_1), \mathbf{not}\, p(b_2). \\
p(b_2) &\leftarrow \mathbf{not}\, p(c_2), \mathbf{not}\, p(b_3). \\
&\;\vdots \\
p(b_n) &\leftarrow \mathbf{not}\, p(c_n), \mathbf{not}\, p(b_{n+1}). \\
p(c_2).
\end{aligned}
$$

that results from instantiating the program from Example 5 using the given base facts and removing obviously true body literals. Note that this grounding operation can be done in linear time w.r.t. $n$. This program can be transformed by $2n$ applications of the transformations *positive reduction* or *negative reduction* into its remainder[1]:

$$
\begin{aligned}
&p(b_1). \\
&p(b_4).\; p(b_6).\; p(b_8).\; \ldots\; p(b_n). \\
&p(c_2).
\end{aligned}
$$

## 5 Intelligent Grounding

The transformations defined in Section 4 are applicable only to ground programs (in (Dix & Stolzenburg, 1998) it is shown how to lift our transformations to first-order disjunctive programs using constraints). To compute the well-founded model of a non-ground program $P$ we could in a first step generate its Herbrand instantiation $ground(P)$. However, this is a very costly operation since many irrelevant rule instances are produced.

*Example 28* (*Complete Instantiation*)
Consider again the program from Example 5. Since it has about $2n$ constants and the first rule has three different variables, its Herbrand instantiation contains $(2n)^3$ ground instances of the first rule. Thus the construction of the Herbrand instantiation would be more expensive than the model computation using the alternating fixpoint procedure (cf. Example 5). Instead we would like to construct a ground instantiation containing only relevant rule instances like that of Example 27 with only $n$ ground instances of the first rule.

In this section we show how for *range-restricted* programs a set of relevant ground rule instances can be derived efficiently. Together with the magic set transformation which we will discuss in Section 8, this technique can be applied to answer queries against programs that are not range-restricted.

The key idea of the proposed grounding algorithm is to compute the set of all possible facts by ignoring all negative body literals during a bottom-up fixpoint iteration. This corresponds to the computation of the set $U_0$ of possible atoms in the alternating fixpoint procedure (cf. Definition 2). But when a rule instance is applied to derive a new atom, not only the head atom but the complete rule instance is stored. We call the ground instances derived this way *conditional facts*.

---

[1] We assume that $n$ is even. Otherwise the second line of the remainder changes slightly.

*Definition 29 (Conditional Fact)*
A *conditional fact* $A \leftarrow \mathcal{C}$ is a ground rule containing possibly both positive and negative delayed literals in the body.

The following immediate consequence operator derives a new conditional fact if for each positive body literal of a ground rule instance there exists at least one matching conditional fact.

*Definition 30 (Immediate Consequences with Conditional Facts)*
Let $P$ be a range-restricted normal program. Given a set $S$ of conditional facts the operator $\bar{T}_P$ computes the following set of conditional facts:

$$\bar{T}_P(S) := \big\{A \leftarrow \mathcal{B} \in ground(P) \mid pos(\mathcal{B}) \subseteq heads(S)\big\}.$$

Since $\bar{T}_P$ is monotonic, its least fixpoint $\text{lfp}(\bar{T}_P)$ exists. Obviously, $\text{lfp}(\bar{T}_P)$ is a subset of $ground(P)$. Because all our rules are range-restricted, the condition $pos(\mathcal{B}) \subseteq heads(S)$ binds all the variables and we never have to resort to a "blind instantiation". More precisely, it is also clear that $heads\big(\text{lfp}(\bar{T}_P)\big) = \text{lfp}(T_{P,\emptyset})$. The operators are very similar, only $T_{P,\emptyset}$ "forgets" the bodies of the applied rule instances.

The following lemma embeds this kind of "intelligent grounding" into our transformation-based approach and shows that it can be seen as an efficient implementation of a part our transformation system.

*Lemma 31 (Intelligent Grounding)*
Let $P$ be a range-restricted normal program. Let $\bar{T}_P$ be defined as above, and let $\mapsto_{LF} := \mapsto_L \cup \mapsto_F$, i.e. the combination of *loop detection* and *failure*. Then we have

$$ground(P) \mapsto^*_{LF} \text{lfp}(\bar{T}_P).$$

*Proof*
See Appendix B.

From Lemma 31 it follows that $\text{lfp}(\bar{T}_P)$ can be used as a starting point for the transformation approach.

*Corollary 32*
Let $P$ be a range-restricted normal program. Then $\text{lfp}(\bar{T}_P) \mapsto^*_X \widehat{P}$ holds.

*Example 33 (Grounding)*
Let $P$ contain the rules and facts from Example 5. Then $\text{lfp}(\bar{T}_P)$ contains

$$
\begin{aligned}
p(a) \quad &\leftarrow \quad t(a, a, b_1), \textbf{not}\, p(a), \textbf{not}\, p(b_1). \\
p(b_1) \quad &\leftarrow \quad t(b_1, c_1, b_2), \textbf{not}\, p(c_1), \textbf{not}\, p(b_2). \\
p(b_2) \quad &\leftarrow \quad t(b_2, c_2, b_3), \textbf{not}\, p(c_2), \textbf{not}\, p(b_3). \\
&\quad\ \ \vdots \\
p(b_n) \quad &\leftarrow \quad t(b_n, c_n, b_{n+1}), \textbf{not}\, p(c_n), \textbf{not}\, p(b_{n+1}). \\
p(c_2) \quad &\leftarrow \quad p_0(c_2).
\end{aligned}
$$

together with the base facts.

The grounding algorithm can be made more efficient by evaluating a program $P$ step by step according to the SCCs of its static dependency graph. Due to the confluence of our transformation system, we can transform the program $P$ SCC by SCC. The following operator enables us to perform the grounding of $P$ also SCC by SCC. These results are not surprising and we refer to (Zukowski *et al.*, 1997) for more details. We will here present only the main result concerning the intelligent grounding.

*Definition 34 (Operator $\hat{T}_{P,R}$)*
Let $P$ be a program and $R$ be a ground program. Given a set $S$ of conditional facts, the operator $\hat{T}_{P,R}$ computes the following set of conditional facts:

$$\hat{T}_{P,R}(S) := \{ A \leftarrow \mathcal{B} \in ground(P) \mid \quad pos(\mathcal{B}) \subseteq heads(R) \cup heads(S) \text{ and} \\ neg(\mathcal{B}) \cap facts(R) = \emptyset \}.$$

If $R$ is empty, we obtain the operator $\bar{T}_P$ of Definition 30 by $\bar{T}_P(S) = \hat{T}_{P,\emptyset}(S)$.

Since this operator is monotonic, its least fixpoint exists. The following definition describes how it can be used for SCC-oriented program grounding and transformation.

*Definition 35 (SCC-Oriented Evaluation)*
Let $P$ be a program. Let $P^{(1)}, \ldots, P^{(m)}$ be a partition of the set of rules of $P$ such that no predicate occurring in $P^{(i)}$ occurs in the heads of $P^{(j)}$ for any $1 \le i < j \le m$. Let the ground programs $R_0, \ldots, R_m$ be defined as follows:

- $R_0 := \emptyset$,
- $R_i$ is the program remainder of $R_{i-1} \cup \text{lfp}(\hat{T}_{P^{(i)}, R_{i-1}})$, for $0 < i \le m$.

The following theorem states that this evaluation scheme is correct and computes the remainder of $P$.

*Theorem 36 (SCC-Oriented Evaluation (Zukowski* et al.*, 1997))*
Let $P$ be a program. Let the sequence $R_0, \ldots, R_m$ be defined as above. Then $R_m$ is the remainder of $P$.

Note that the bottom-up fixpoint computation of the grounding as described above performs an implicit *loop detection* for each SCC. As a consequence, for a large program class no explicit loop detection has to be applied.

*Lemma 37 (Avoiding Loop Detection (Zukowski* et al.*, 1997))*
Let $P$ be a program. Let $P$ contain no predicate which depends on itself both, positively and negatively (i.e. through negation). Then during the SCC-oriented grounding and transformation of $P$ as defined in Definition 35 no explicit application of *loop detection* is necessary.

Further optimization can be obtained by immediately removing obviously true body literals from the conditional facts already during the fixpoint computation. This way normal (unconditional) facts are produced whenever the truth value of all body literals is known whereas conditional facts are produced only if the predicate of a negative body literal is in the same SCC as the head predicate or a body literal depends already on a conditional fact.

*Example 38* (*Optimized Grounding*)
Consider again the program from Example 5. If during the grounding algorithm body literals that are known to be true are immediately removed, the simplified ground program from Example 27 is constructed, together with the base facts. Only literals with the recursive predicate $p$ appear in conditions.

Note that for *predicate stratified programs* (Przymusinski, 1989a) the SCC-oriented grounding with immediate removal of true body literals always computes the well-founded model of the program directly. Since the truth values of all negative body literals are defined in lower SCCs, there is no need to produce any conditional facts. Thus the optimized grounding algorithm is a smooth generalization of the standard bottom-up fixpoint computation. Stratified parts of the program are evaluated as usual. Conditional facts are only generated for negative recursion.

## 6 Relation to Alternating Fixpoint

The computation of the set of all possibly true facts that is needed for the *loop detection* $\mapsto_L$ transformation (cf. Definition 17) is very similar to the computation of possible facts performed in each iteration step of the alternating fixpoint procedure (c.f. Definition 2). There is an even closer correspondence as we will show in this section.

Our main result is that a particular ordering of our transformations corresponds exactly to the alternating fixpoint procedure. To characterize the computation performed by the alternating fixpoint procedure we define a sequence $(\bar{K}_i, \bar{U}_i)_{i \geq 0}$ where $\bar{K}_i$ and $\bar{U}_i$ are constructed by the program transformations defined in this paper and that is closely related to the sequence $(K_i, U_i)_{i \geq 0}$ computed by the alternating fixpoint procedure (cf. Definition 2).

*Definition 39* (*Alternating Fixpoint by Program Transformations*)
Let $P$ be a normal logic program. We define the sequence $(\bar{K}_i, \bar{U}_i)_{i \geq 0}$ of sets $\bar{K}_i$ and $\bar{U}_i$ of conditional facts as follows:

$$
\begin{aligned}
\bar{K}_0 \quad & \text{is the normal form of } ground(P) \text{ w.r.t. } \mapsto_{LFS}, \\
& \text{i.e., } ground(P) \mapsto_{LF}^* \mathrm{lfp}(\bar{T}_P) \mapsto_S^* \bar{K}_0, \\
\bar{U}_0 \quad & \text{is the normal form of } \bar{K}_0 \text{ w.r.t. } \mapsto_{NLF}, \\
i > 0 : \quad \bar{K}_i \quad & \text{is the normal form of } \bar{U}_{i-1} \text{ w.r.t. } \mapsto_{PS}, \\
i > 0 : \quad \bar{U}_i \quad & \text{is the normal form of } \bar{K}_i \text{ w.r.t. } \mapsto_{NLF}.
\end{aligned}
$$

The computation terminates when the sequence becomes stationary, i.e., when a fixpoint is reached in the sense that

$$(\bar{K}_j, \bar{U}_j) = (\bar{K}_{j+1}, \bar{U}_{j+1}).$$

*Lemma 40* (*Confluence of $\mapsto_{PS}$ and $\mapsto_{NLF}$*)

In Definition 39, the normal form of $\bar{U}_i$, $i \geq 0$, w.r.t. $\mapsto_{PS}$ and the normal form of $\bar{K}_i$, $i \geq 0$, w.r.t. $\mapsto_{NLF}$ are uniquely determined.

*Proof*
See Appendix C.

*Remark 41*
Due to Corollary 25 and Lemma 31 the sequence $(\bar{K}_i, \bar{U}_i)_{i \geq 0}$ of Definition 39 reaches a fixpoint $(\bar{K}_j, \bar{U}_j)$ and computes the program remainder $\widehat{P} = \bar{K}_j = \bar{U}_j$.

The computation of the sets $K_i$ of true facts by the alternating fixpoint procedure corresponds to the application of *success* and *positive reduction* on the transformation side. The computation of the sets $U_i$ of possible facts corresponds to deleting conditional facts by applying *negative reduction*, *loop detection*, and *failure*. The relation between the alternating fixpoint computation and the sequence of program transformations defined above is described more precisely by the following theorem.

*Theorem 42 (Relation between AFP and Transformations)*
Let $P$ be a normal logic program. Let $(K_i, U_i)_{i \geq 0}$ and $(\bar{K}_i, \bar{U}_i)_{i \geq 0}$ be the sequences given by Definitions 2 and 39, respectively. Then the following holds: $K_i = facts(\bar{K}_i)$ and $U_i = heads(\bar{U}_i)$.

*Proof*
See Appendix C.

Note that the last theorem states the correctness of the transformation method only for one specific strategy. Thus it does no restate Theorem 23 which is more general and shows the correctness for any transformation strategy.

Theorem 42 shows that the program transformation approach, if a transformation strategy like that given in Definition 39 is applied, will never do more work than the alternating fixpoint procedure. Although the number of conditional facts in the sets $\bar{K}_i$ and $\bar{U}_i$ may be greater than the number of unconditional facts in the corresponding sets $K_i$ and $U_i$, the same number of computation steps is needed to derive them. The existence of several conditional facts having the same head shows that an atom can be derived by more than one rule application. Also the alternating fixpoint computation derives facts more than once, but performs a duplicate elimination afterwards. When a new fact is generated by the transformation approach by more than one transformation application to reduce all body literals of a rule instance, also the alternating fixpoint computation has to consider all body literals before the head of an rule instance can be derived.

However, while having a (sometimes) better time complexity, the transformation approach is expected to have a space complexity worse than the alternating fixpoint approach.

As proposed in (Kemp *et al.*, 1995), the sets $K_i$ do not have to be recomputed in each iteration step, but can be maintained in terms of increments. This optimization is automatically performed by the transformation approach, since conditional facts that were transformed into facts by the deletion of their body literals are never changed afterwards.

Note that the results of this section show that the alternating fixpoint procedure can be seen as one particular implementation of our approach (given by choosing a fixed ordering of applying the transformations). However, by choosing another ordering that delays the application of *loop detection* until no other transformation is applicable, the computation scheme will be much more efficient for many programs. This has already been demonstrated in Examples 5 and 27. We will further illustrate this point in the next section.

## 7 Regular Strategy Expressions

In Section 6 we have seen that we can choose the evaluation strategy just by rearranging the order of transformation application. In order to specify a transformation order we introduce the following regular expression syntax.

*Definition 43 (Regular Strategy Expression)*
A *regular strategy expression* is defined inductively as follows:

- The single letters $P$, $S$, $N$, $F$, and $L$ are *atomic* regular strategy expressions.
- If $e_1$ and $e_2$ are regular strategy expressions, then $e_1 e_2$ is a regular strategy expression.
- If $e$ is a regular strategy expression, then $(e)^*$ is a regular strategy expression.

Let $P$ be a ground program. Let $e$ be a regular strategy expression. $P'$ is the result of the *application of e to P*, $P \overset{e}{\mapsto} P'$, if the following conditions hold:

- If $e$ is an *atomic* regular strategy expression, i.e., there is a ground transformation $\mapsto_e$, then $P'$ is the normal form of $P$ w.r.t. $\mapsto_e$.
- If $e = e_1 e_2$, then there exists a ground program $P''$ such that $P \overset{e_1}{\mapsto} P''$ and $P'' \overset{e_2}{\mapsto} P'$ hold.
- If $e = (c)^*$, where $c$ may be a *complex*, non-atomic strategy expression, then $P'$ results from iteratively applying $c$ to $P$ until no changes occur.

*Lemma 44 (Unique Result)*
Let $P$ be a ground program. Let $e$ be a regular strategy expression. Then the result of applying $e$ to $P$ is uniquely determined, i.e., there is exactly one $P'$ such that $P \overset{e}{\mapsto} P'$ holds.

*Proof*
The proof is by induction over the structure of the regular strategy expression $e$ and is based on the fact that the normal form of each base transformation is uniquely determined.    □

If we compare strategies, we assume tacitly that the grounding algorithm of Section 5 is prepended such that the execution of the strategies always starts from the set of relevant ground rule instances. This is fair, since all bottom-up methods have to do the grounding operations, even if they are performed on the fly later during the computation.

*Example 45 (Fitting Expression)*
Following Theorem 15, the least fixpoint of the Fitting operator corresponds to the normal form $fitt(P)$ of the rewriting system $\mapsto_{PSNF}$. Thus the computation of $fitt(P)$ for a given program $P$ can be expressed by the regular strategy expression $(PSNF)^*$. Since a linear time algorithm for the Fitting operator applied to ground programs is given in (Niemelä & Simons, 1996), for a ground program $P$ the transformation $P \overset{(PSNF)^*}{\mapsto} fitt(P)$ can be executed in time $O(n)$ where $n$ is the size of $P$. Note that here the order of the four transformations has no influence on the time complexity.

*Example 46 (AFP Expression)*
Following Definition 39 and Theorem 42 one iteration of the alternating fixpoint procedure can be described by the expressions $PS$ for the computation of $\bar{K}_i$ and $NLF$ for the computation of $\bar{U}_i$. Thus, the model computation of the alternating fixpoint procedure can be described by the regular strategy expression $(PSNLF)^*$. Since each application of the *loop detection* needs $O(n)$ time, the execution time of the transformation $P \overset{(PSNLF)^*}{\mapsto} \widehat{P}$ is in $O(n^2)$ where $n$ is the size of the given ground program $P$. Note that the expression $PS$ is equivalent to the expression $(PS)^*$, and $NLF$ is equivalent to $(NLF)^*$, i.e., the loop is always executed exactly once. This becomes evident in the proof of Lemma 40.

In the expression $(PSNLF)^*$ of Example 46 the expensive *loop detection* is applied in every iteration. In the following optimized strategy *loop detection* is removed from the inner loop and is only applied if no other transformation is applicable.

*Example 47 (Optimized Remainder Expression)*
A more efficient strategy to compute the remainder of a given program is expressed by the regular strategy expression $((PSNF)^*L)^*$. For programs without positive loops, the transformation $P \overset{((PSNF)^*L)^*}{\mapsto} \widehat{P}$ can be executed in linear time, since the inner loop $(PSNF)^*$ computes the remainder and *loop detection* is not applicable. For programs with positive loops $O(n)$ iterations of the outer loop are necessary yielding again an execution time in $O(n^2)$ where $n$ is size of the given ground program.

*Example 48 (Experimental Results)*
Consider the program shown in Example 5 and its ground instantiation from Example 27. Figure 1 illustrates execution times of the AFP strategy from Example 46 and the optimized remainder strategy from Example 47 for different program sizes $n$ ranging from 100 to 1000. The quadratic and linear character of the two strategies, respectively, is obvious.

The tests have been run using our prototypical implementation in Java on a PC with an Intel Pentium III processor with 450 MHz and 256 MB of main memory, MS Windows NT 4.0 operating system and Sun JDK $1.3\beta$ java virtual machine. The same test running on a Sun Sparc Station with Sun Solaris operating system yields comparable results.

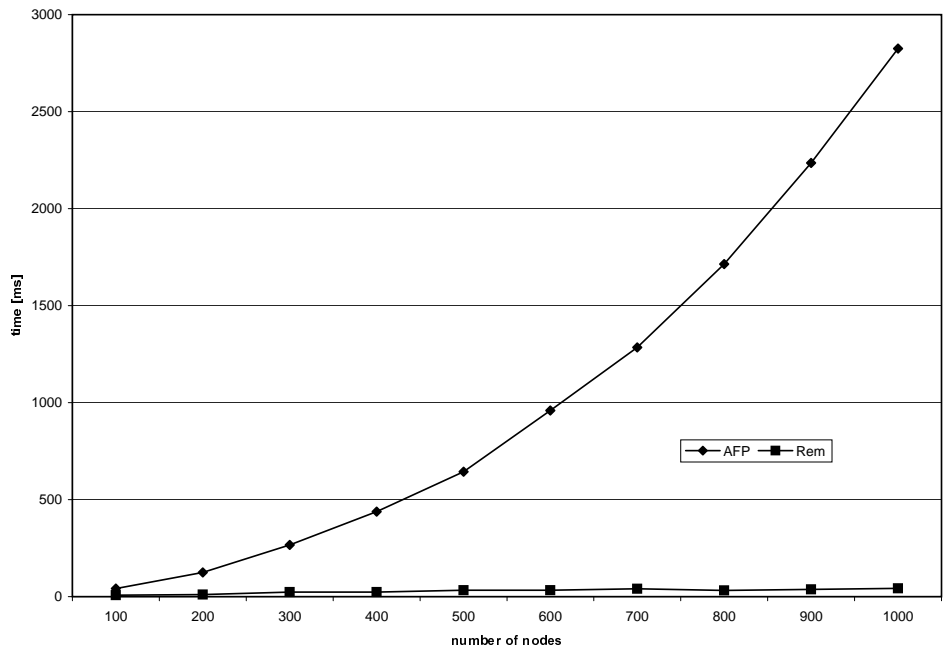

Fig. 1. Experimental Results

## 8  Magic Transformation

It is well-known that the magic set transformation (Beeri & Ramakrishnan, 1991) causes problems in the context of the well-founded semantics. For a given query the well-founded model of the magic set transformed program may not agree with the well-founded model of the original program on the query, if magic facts get undefined truth values in the model of the transformed program. There are two prominent extensions of the alternating fixpoint procedure that solve this problem by considering undefined magic sets to be true at different steps of the computation. The first approach is the *well-founded magic set* method.

*Definition 49* (*Well-Founded Magic Set (Kemp* et al., *1995)*)
For a given normal program $P$, any sideways information passing (sip) strategy $S$, and a query $Q$, the *well-founded magic set* method (Kemp *et al.*, 1995) determines the magic set transformed program $MP = Magic(P, S, Q)$ of $P$ w.r.t. $S$ and $Q$. Then the set $M$ of true or undefined magic facts in the well-founded model $W_{MP}^*$ is computed using the alternating fixpoint procedure. Finally the well-founded model of $WFMagic(P, S, Q) := MP \cup M$ is computed, again using the alternating fixpoint procedure.

We do not discuss the details of different sideways information passing strategies, since they are not relevant for our results, and assume that always standard left to right binding propagation is applied. For details we refer the reader to (Beeri & Ramakrishnan, 1991; Kemp *et al.*, 1995).

The correctness of the method described in Definition 49 is stated by the following theorem.

**Theorem 50** (*Well-Founded Magic Set (Kemp* et al., *1995)*)
Let $P$ be a normal logic program. Let $S$ be any sip strategy (cf. Definition 49). Let $Q$ be a query. Let $PM := WFMagic(P, S, Q)$. Then the well-founded models of $PM$ and $P$ agree on $Q$.

To use this result in our approach we extend our transformation system by the following transformation that allows us to use undefined magic facts as if they were true.

**Definition 51** (*Magic Reduction*)
Let $P_1$ and $P_2$ be ground programs. $P_2$ results from $P_1$ by *magic reduction* ($P_1 \mapsto_M P_2$) iff there is a rule $A \leftarrow \mathcal{B}$ in $P_1$ with head predicate $p$ and a positive literal $B \in \mathcal{B}$ with predicate *magic_p* such that $B \in heads(P_1)$ and $P_2 = (P_1 - \{A \leftarrow \mathcal{B}\}) \cup \{A \leftarrow (\mathcal{B} - \{B\})\}$.

Our final transformation system is defined as follows.

**Definition 52** (*MX-Transformation*)
Let $\mapsto_{MX}$ denote the extended rewriting system:

$$\mapsto_{MX} := \mapsto_X \cup \mapsto_M.$$

Note that this new transformation system is not confluent any more! Depending on the application of the *magic reduction* transformation different strategies for handling undefined magic facts can be realized. However, the following theorem states that any strategy based on the $MX$-transformation computes the correct answer for the given query.

**Theorem 53** (*Correctness of the MX-Transformation*)
Let $P$ be a normal logic program. Let $S$ be any sip strategy. Let $Q$ be a query. Let $MP = Magic(P, S, Q)$ be as defined above. Let $P'$ be any normal form of $MP$ w.r.t. $\mapsto_{MX}$. Then the well-founded models of $P'$ and $P$ agree on $Q$.

*Proof*
This result is based on the fact that all magic facts that are undefined in the well-founded model if $MP$ can be considered true by applications of the *magic reduction* transformation. In this case the correctness follows from Theorem 50. But it can be the case that a conditional magic fact gets deleted before all conditions consisting of its head have been removed by applications of *magic reduction*. In this case the magic fact has become false because at least one of the body literals in each of its conditional facts have become false. But then also the rule instance for that the magic fact had been generated can be deleted, because its body contains all literals that the magic fact depends on, due the definition of the magic set transformation. Thus, the deleted magic fact is not needed to answer the given query. As a consequence, all magic facts needed to answer the given query are not deleted and can be considered to be true by applications of the *magic reduction* transformation.    □

We now extend our regular strategy expression syntax by the new transformation.

*Definition 54 (Regular Strategy Expression (cont.))*
We extend Definition 43 in the following way. The single letter $M$ is also an atomic regular strategy expression.

According to Definition 43, the expression $M$ denotes the computation of the normal form w.r.t. $\mapsto_M$, i.e. the removal of all true or undefined magic filters from rules with heads with non-magic predicates. This allows us to embed the well-founded magic set method as a special case of our transformation method.

*Definition 55 (Well-Founded Magic Sets Strategy)*
The well-founded magic sets regular strategy expression is defined by $(PSNLF)^*M(PSNLF)^*$.

*Lemma 56*
Let $MP$ be a magic set transformed ground program. The execution of the regular strategy expression $(PSNLF)^*M(PSNLF)^*$ applied to $MP$ corresponds to the computation of the well-founded magic sets method of Definition 49.

*Proof*
This result follows from Theorem 42, Example 46, and Definitions 51 54. □

It is obvious that the well-founded magic set strategy can be optimized by delaying the *loop detection* transformation.

*Definition 57 (Well-Founded Remainder)*
The *well-founded remainder* regular strategy expression is defined by $((PSNF)^*L)^*M((PSNF)^*L)^*$.

*Lemma 58*
Let $MP$ be a magic set transformed ground program. The execution of the regular strategy expression $((PSNF)^*L)^*M((PSNF)^*L)^*$ applied to $MP$ is guaranteed to need no more work than the transformation-based computation described by the expression $(PSNLF)^*M(PSNLF)^*$. Furthermore, it is more efficient for many programs.

*Proof*
This statement follows directly from Theorem 42. □

One disadvantage of the two strategies in Lemma 58 is that the application of the *magic reduction* transformation all undefined magic facts are considered to be true and removed from all bodies of non-magic rules. This can cause unnecessary computations since as a consequence of making some magic facts true it can happen that other magic facts become false, i.e., the computation of the corresponding subquery is not needed to answer the query. This is the motivation for another approach. The *magic alternating fixpoint* method uses a modified version of the alternating fixpoint procedure where in every computation of true facts all magic facts that were undefined in the preceding computation of true or possible facts are considered true.

*Definition 59 (Magic Alternating Fixpoint (Morishita, 1996))*
Let $P$ be a normal logic program. Let $S$ be any sip strategy (cf. Definition 49). Let $Q$ be a query. Let $MP = Magic(P, S, Q)$ be the magic transformed program of $P$ w.r.t. $S$ and $Q$. Let $MP^+$ denote the subprogram consisting of the definite rules of $MP$. The *magic alternating fixpoint procedure* computes the sequence $(K_i, U_i)_{i \geq 0}$ defined as follows:

$$
\begin{array}{rcl}
K_0 & := & \mathrm{lfp}(T_{MP^+}) \\
U_0 & := & \mathrm{lfp}(T_{MP, K_0}) \\
i > 0: \quad K_i & := & \mathrm{lfp}(T_{MP \cup magic\_heads(U_{i-1}), U_{i-1}}) \\
i > 0: \quad U_i & := & \mathrm{lfp}(T_{MP, K_i}).
\end{array}
$$

where $magic\_heads(U_{i-1})$ denotes the set of magic atoms in $U_{i-1}$.

The correctness of this method is stated by the following theorem.

*Theorem 60 (Magic Alternating Fixpoint (Morishita, 1996))*
Let $P$ be a normal logic program. Let $S$ be any sip strategy (cf. Definition 49). Let $Q$ be a query. Let the sequence $(K_i, U_i)_{i \geq 0}$ be defined as above. Then the computation gets stationary, i.e., there is an $j$ such that $K_j = K_{j+1}$ and $U_j = U_{j+1}$. Furthermore, the corresponding interpretation $I = K_j \cup \sim(BASE(P) - U_j)$ and the well-founded model $W_P^*$ of $P$ agree on $Q$.

This approach dynamically changes the set of true magic atoms. The advantage of this method is stated by the following theorem.

*Theorem 61 (Comparison (Kemp et al., 1995; Morishita, 1996))*
Let $P$ be a normal logic program. Let $Q$ be a query. The magic alternating fixpoint procedure is guaranteed to terminate after no more iterations than the well-founded magic sets approach, and the set of magic facts in the final set $U_j$ of the magic alternating fixpoint procedure is a subset of the set $M$ of true or undefined magic sets in the well-founded magic sets approach.

In the magic alternating fixpoint procedure (Morishita, 1996) undefined magic facts are considered to be true only if this enables the derivation of new facts. This can be described by a special case of our *magic reduction* transformation. The *restricted magic reduction* transformation allows to remove a magic filter with undefined truth value from a non-magic rule only if it is the only body literal, i.e., the application transforms the rule into a fact.

*Definition 62 (Restricted Magic Reduction)*
from $P_1$ by *restricted magic reduction* $(P_1 \mapsto_R P_2)$ iff there is a rule $A \leftarrow \{B\}$ in $P_1$ with head predicate $p$ such that the predicate of $B$ is $magic\_p$, $B \in heads(P_1)$ and $P_2 = (P_1 - \{A \leftarrow \mathcal{B}\}) \cup \{A\}$.

It is easy to see that this *restricted magic reduction* transformation is sufficient to evaluate magic set transformed programs.

*Lemma 63 (Correctness of Restricted MX-Transformation)*

Let $MP$ be a magic set transformed ground program. Let $P'$ a normal form of $MP$ w.r.t. the rewriting system $\mapsto_X \cup \mapsto_R$. Let $P''$ be a normal form of $P'$ w.r.t. $\mapsto_{MX}$. Then $known_{BASE(P)}(P') = known_{BASE(P)}(P'')$ holds.

*Proof*
The only transformation applicable to $P'$ can be $\mapsto_M$ but without producing new facts. Thus no other transformation will become applicable, no new facts can be produced, and no rules can be deleted from $P'$. $\quad\square$

The following definition allows us to use the *restricted magic reduction* transformation in regular strategy expressions.

*Definition 64 (Regular Strategy Expression (cont.))*
We extend the Definitions 43 and 54 in the following way. The single letter $R$ is also an atomic regular strategy expression.

Now we can express the magic alternating fixpoint procedure as a special case of our rewriting system $\mapsto_{MX}$.

*Definition 65 (Magic Alternating Fixpoint Strategy)*
The Magic Alternating Fixpoint regular strategy expression is defined by $(P(SR)^*NLF)^*$.

*Lemma 66*
Let $MP$ be a magic set transformed ground program. The execution of the regular strategy expression $(P(SR)^*NLF)^*$ applied to $MP$ corresponds to the computation of the magic alternating fixpoint procedure of Definition 59.

*Proof*
The magic alternating fixpoint method modifies the alternating fixpoint procedure such that during the computation of true atoms undefined magic filters are considered to be true. This means that we have to replace the expression $S$ in the description of the alternating fixpoint procedure by the expression $(SR)^*$ that also computes the normal form of the *success* transformation but additionally is allowed to remove undefined magic filters from non-magic rules if this leads directly to the derivation of facts. $\quad\square$

*Remark 67 (Monotonicity)*
Note that the sequence of sets $K_i$ of true facts is not monotonically increasing in the magic alternating fixpoint procedure, because some magic facts and facts depending on them that have been considered true once may possibly not be recomputed in a later iteration step. However, in the magic alternating fixpoint transformation, magic conditional facts may be deleted, but non-magic facts once derived as unconditional facts are never deleted afterwards. Thus, our transformational counterpart of the magic alternating fixpoint procedure has the monotonicity property that is missing in the original approach of MORISHITA (Morishita, 1996).

Examining the magic alternating fixpoint strategy expression $((P(SR)^*NLF)^*$ there are two possible optimizations. The more obvious optimization is to delay the application of *loop detection*. The second optimization concerns the time when undefined magic facts are to be considered true. In (Kemp *et al.*, 1995) it is argued that using magic facts too early may cause unnecessary computations since the magic undefined magic facts may become false later. Therefore we delay also the application of the *restricted magic reduction* in the following strategy.

*Definition 68 (Magic Remainder)*
The Magic Remainder regular strategy expression is defined by $(((PSNF)^*R)^*L)^*$.

*Lemma 69*
Let $MP$ be a magic set transformed ground program. The execution of the regular strategy expression $((PSNF)^*R)^*L)^*$ applied to $MP$ is guaranteed to need no more work than the transformation-based computation described by the expression $(P(SR) * NLF)^*$ and is more efficient for many programs.

*Proof*
The expression $((PSNF)^*R)^*$ can be executed in linear time w.r.t. the given ground program. For the expression $(PSNF)^*$ this has been discussed before. For the *restricted magic reduction* each rule has to be checked at most once, i.e., if all body literals except the magic filter have been removed. This operation adds extra costs that are also linear in the size of $MP$. The only expensive operation is the *loop detection* transformation which is removed from the inner loop and is applied only if no other transformation is applicable. An example for a program that is evaluated more efficiently by the transformation approach is shown in Example 71. □

The following proposition is the main result of this section.

*Proposition 70 (Magic Remainder vs. Well-Founded Magic Sets)*
The computation corresponding to the magic remainder strategy is guaranteed to need no more work than the computation performed by the well-founded magic sets method or the magic alternating fixpoint and is more efficient for many programs.

*Proof*
This result follows directly from the Lemmata 56, 58, 66, and 69. □

The statement of this proposition is illustrated by the following example.

*Example 71*
Consider the following program $MP$

$$
\begin{aligned}
p(X) \quad &\leftarrow \quad m\_p(X), t(X,Y,Z), \mathbf{not}\, p(Y), \mathbf{not}\, p(Z). \\
p(X) \quad &\leftarrow \quad m\_p(X), p_0(X). \\
m\_p(a). \quad & \\
m\_p(Y) \quad &\leftarrow \quad m\_p(X), t(X,Y,Z). \\
m\_p(Z) \quad &\leftarrow \quad m\_p(X), t(X,Y,Z), p(Y).
\end{aligned}
$$

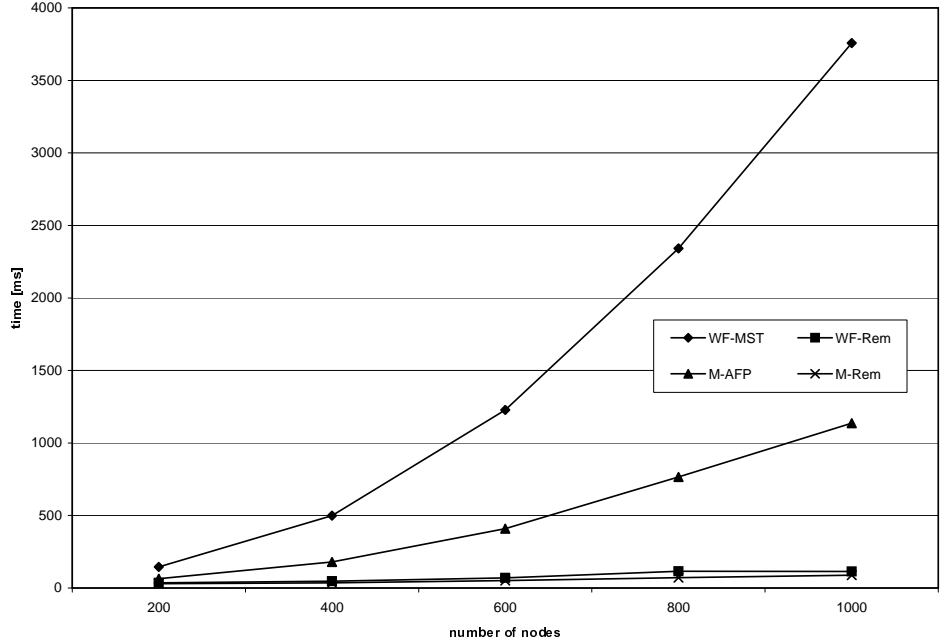

Fig. 2. Comparison without positive Loops

that results from the program $P$ from Example 5 by the magic set transformation for the query $p(a)$ together with the base facts

$$p_0(c_{\frac{n}{4}}), t(a, a, b), t(b_1, c_1, b_2), t(b_2, c_2, b_3), \ldots, t(b_n, c_n, b_{n+1}).$$

Figure 2 illustrates the execution times of the four strategies of Definitions 55 (WF-MST), 57 (WF-Rem), 65 (M-AFP) and 68 (M-Rem) for different program sizes $n$. It can be seen that the WF-MST strategy makes all magic facts true and needs quadratic time to compute the complete well-founded model of the original program. The M-AFP strategy uses only the relevant magic facts and computes only the facts from $p(b_1)$ to $p(b_{\frac{n}{4}})$, but also needs quadratic time. The strategies WF-Rem and M-Rem compute the result in linear time, since the program does not contain positive loops and thus the *loop detection* is not needed. Consider now the same program but with an additional rule $p(X) \leftarrow p(X)$. Now every $p$-fact depends positively on itself and a linear number of applications of *loop detection* is needed to compute the result. Figure 3 illustrates the execution times of the four strategies for the modified program. It can be seen that the execution times of the strategies M-AFP and M-Rem coincide and are less than the execution times of the strategies WF-MST and WF-Rem, which also coincide. The delay of the *loop detection* has no effect for this program.

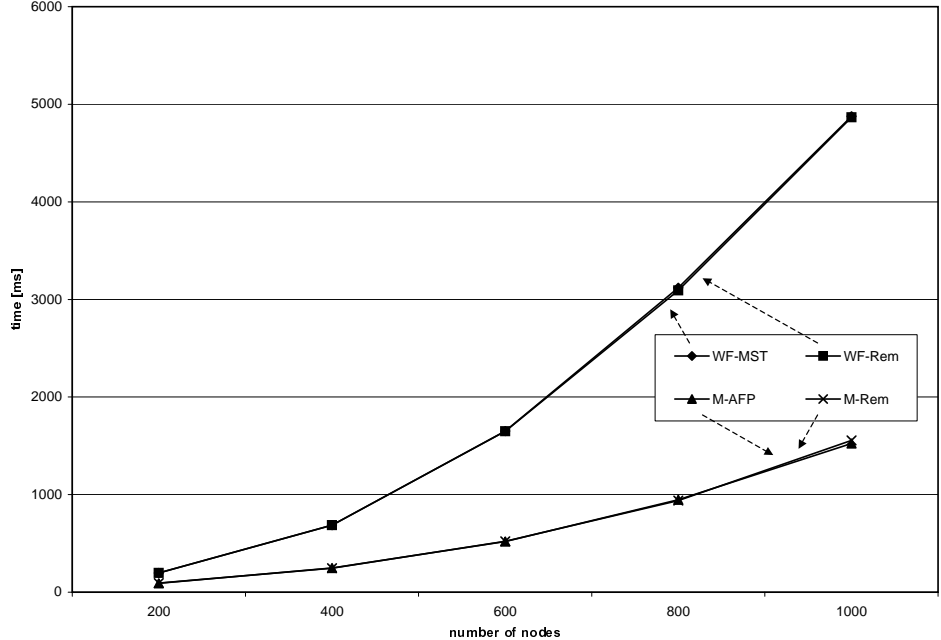

Fig. 3. Comparison with positive Loops

## 9 Discussion

### *Residual Program Approach*

The concept of conditional facts and the transformation of ground programshave
been introduced and investigated in (Brass & Dix, 1994; Brass & Dix, 1997; Brass
& Dix, 1998; Brass & Dix, 1999). Disjunctive ground programs are transformed into
the so called *residual program* from which several semantics can be derived. Whereas
the transformations *positive* and *negative reduction* are defined as in our approach,
they allow full *unfolding* instead of only the special cases *success* and *failure*. As
a consequence, the residual program can grow to a size that is *exponential* in the
size of the original program. As a replacement for their *unfolding* and *deletion of
tautologies* transformations we introduced the *loop detection* to remove positive
loops. This way, the size of the program is reduced in every transformation step
yielding a linear space complexity w.r.t. the size of the original program.

### *Alternating Fixpoint Procedure*

We have presented a precise comparison to the *alternating fixpoint procedure*
(Van Gelder, 1993) in Section 6. In Theorem 42 we showed that one specific ordering
of our transformations corresponds exactly to the computation of the alternating
fixpoint procedure. We proposed an evaluation strategy that is guaranteed to need
no more work than the alternating fixpoint procedure but that needs linear time to compute the model of many programs the alternating fixpoint approach needs quadratic time for.

The key idea for this improvement is to avoid the repeated computations of possible facts of the alternating fixpoint procedure. Instead, our grounding algorithm derives *conditional facts*, i.e., ground rule instances, that can be reduced efficiently. The expensive *loop detection* transformation is applied only if necessary. Of course, this improvement of the time complexity results in a possibly increased space complexity.

It is well known that the alternating fixpoint procedure needs quadratic time even for programs without positive loops, e.g., if there is a sequence of negative dependencies between ground atoms as in Example 5. In Lemma 37 we showed that our approach, in combination with SCC-oriented grounding, needs to apply loop detection only if any predicate occurring in a positive loop depends negatively on itself. For all other programs the well-founded model can be computed in linear time w.r.t. the size of the generated ground programs by applying the four base transformations corresponding to the Fitting operator (cf. Theorem 15 and Example 45).

### smodels

The optimized alternating fixpoint strategy of Example 47 corresponds exactly to the well-founded model computation of the *smodels* approach (Niemelä & Simons, 1996). The grounding algorithm of smodels also derives relevant ground instances of the given program rules, i.e., it generates conditional facts. In (Niemelä & Simons, 1996) a linear-time algorithm for the Fitting operator is described that corresponds exactly to the computation of the normal form w.r.t. our $\mapsto_{PSNF}$ transformation. In (Niemelä & Simons, 1996) also a linear-time algorithm for the *loop detection* based on the evaluation method for propositional programs of (Dowling & Gallier, 1984) is presented which we could adapt to our purposes. Thus the well-founded model computation of smodels can be considered as an efficient implementation of a specific strategy of our approach. However, it is not goal-directed and applicable only to range-restricted programs. As opposed to smodels, our approach is applicable to magic set transformed programs and thus allows a goal-directed evaluation of a large class of non range-restricted programs.

### *Magic Set Transformation*

The alternating fixpoint procedure has been extended to magic set transformed programs (Beeri & Ramakrishnan, 1991) by several researchers. The *well-founded magic sets* methods (Kemp *et al.*, 1995) and the *magic alternating fixpoint procedure* (Morishita, 1996) treat magic set transformed programs by considering undefined magic facts to be true at different stages of the computation. The *well-founded ordered search* algorithm (Stuckey & Sudarshan, 1997) is an extension of the *ordered search* method (Ramakrishnan *et al.*, 1992), that applies the alternating fixpoint procedure whenever a negative cyclic dependency is detected.

We have shown in Lemmata 56 and 66 that the first two methods are instances of our transformation approach. We proposed evaluation strategies (cf. Definitions 57 and 68) that combined the correct treatment of magic set transformed programs with the efficient reduction of conditional facts of the smodels approach. Thus we presented an algorithm that combines the advantages of several methods, i.e. the goal-directed and set-oriented bottom-up computation without the unnecessary re-computations of possible facts.

The well-founded ordered search method is not an instance of our transformation framework since it relies on a more complex bottom-up computation that keeps track of all dependencies between ground atoms. However, when a negative cyclic dependency if detected, an algorithm based on the alternating fixpoint us used to resolve them. At this point, our more efficient remainder computation based on conditional facts instead of simple atoms may be applied to improve the overall performance of the well-founded ordered search method.

### *SLG*

The most prominent top-down methods for the computation of the well-founded semantics are the *tabulated resolution* of Bol and Degerstedt (Bol & Degerstedt, 1998) and the *SLG resolution* (Chen & Warren, 1996). Our approach is closely related to the SLG resolution, i.e., the transformation system $\mapsto_X$ can be seen as a bottom-up variant of the SLG resolution restricted to the model computation of ground programs. Together with the magic set transformation and the intelligent grounding algorithm, our approach allows also query answering against some non range-restricted programs. Thus for some applications, it may serve as an appropriate bottom-up counterpart of the top-down SLG resolution.

One major advantage of the SLG resolution is that the activation of subgoals and the instantiation of rules can be interleaved with the reduction of true or false body literals. SLG tries to evaluate completely one subgoal before the next body literal of a rule is activated. If one subgoal fails, the evaluation of the corresponding rule is terminated and the remaining subgoals are not instantiated. In our approach, each SCC has to be grounded completely before the transformation is applied to reduce true or false body literals.

On the other hand, if large base relations are supplied, e.g. by external databases, a set-oriented evaluation is desired instead of a tuple-at-a-time strategy. For the SLG resolution an alternative scheduling strategy has been implemented (Freire *et al.*, 1997) that realizes a breadth-first set-at-a-time tabling strategy that has been proven to be iteration-equivalent to a form of semi-naive magic evaluation. We propose the evaluation in two phases, a bottom-up computation using optimized relational techniques, followed by a reduction phase for the case that any conditional facts have been derived. Our approach supports a set-oriented evaluation and allows mixed top-down and bottom-up strategies. As we have verified in our prototype implementation, it is straightforward to extend the classical relational algebra operations and the differential fixpoint iteration (Balbin & Ramamohanarao, 1986) to generate conditional facts in the presence of negative body literals.

Another advantage of our transformation system is its *simplicity*. Most computation methods for the well-founded semantics (except the alternating fixpoint procedure) need sophisticated data structures or algorithms already to guarantee a correct evaluation. Our approach is easy to understand, thus appropriate for educational purposes, and easy to implement, as we have shown by our prototypical implementation. Different optimizations can be applied, and of course have to be applied in order to get a competitive real life system, but these optimizations are not predetermined by the method. Furthermore, extensions to the transformation system, e.g. to compute a variant of the well-founded semantics, are straightforward.

## 10 Conclusion

We have presented an algorithm for computing the well-founded semantics of function-free programs which is based on the set of elementary program transformations studied by BRASS and DIX (Brass & Dix, 1994; Brass & Dix, 1997; Brass & Dix, 1998; Brass & Dix, 1999). The execution time of our algorithm is polynomial in the size of the EDB. If all positive body literals in the program remainder were replaced (and tautologies were deleted), we would get the residual program plus some non-minimal conditional facts. So the program remainder can be seen as an efficient encoding of the residual program. It is equivalent to the original program under a wide range of semantics. In fact, our rewriting system $\mapsto_X$ is even weaker than the transformations used in (Brass & Dix, 1994; Brass & Dix, 1999) because full *unfolding* is not needed to compute the program remainder.

We have shown the flexibility of our method by introducing *regular strategy expressions*. In particular, we have presented strategies corresponding to the computation of the least fixpoint of the Fitting operator, to the alternating fixpoint approach (Van Gelder, 1989; Van Gelder, 1993), to the well-founded magic sets approach of (Kemp *et al.*, 1995) and to the magic alternating fixpoint approach of (Morishita, 1996). By using the confluence of our calculi or choosing better strategies, we are never doing more work than these methods: as shown by experimental results, there are many examples where our method is superior.

In fact our approach also extends to magic sets methods and avoids the recomputations of many facts.

We have shown that many bottom-up computation methods that are based on the alternating fixpoint procedure can profit directly from our results. Using conditional facts in the well-founded magic set method (Kemp *et al.*, 1995), the magic alternating fixpoint procedure (Morishita, 1996), and the well-founded ordered search algorithm (Stuckey & Sudarshan, 1997), unnecessary computations can be avoided in all three approaches. The combination of the goal-directed search strategy of ordered search together with our concepts of conditional facts to store and later reduce possible facts yields a method that may be a competitive bottom-up counterpart of the top-down method SLG (Chen & Warren, 1996).

## Acknowledgements

We would like to thank two referees, the comments of which helped to put this paper in its current form. Their pointers to related literature and their critical remarks to various issues in previous versions were invaluable and helped to extend and to improve this article a lot.

# A  Proofs of Section 4

*Definition 72 (Allowed Transformation)*
A semantics $\mathcal{S}$ allows a transformation $\mapsto$ iff $\mathcal{S}(P_1) = \mathcal{S}(P_2)$ for all $P_1$ and $P_2$ with $P_1 \mapsto P_2$.

*Definition 73 (Deletion of Tautologies)*
Let $P_1$ and $P_2$ be ground programs. Program $P_2$ results from program $P_1$ by *deletion*

*of tautologies* $(P_1 \ \mapsto_T \ P_2)$ iff there is $A \leftarrow \mathcal{B} \ \in \ P_1$ such that $A \in \mathcal{B}$ and $P_2 = P_1 - \{A \leftarrow \mathcal{B}\}$.

*Definition 74* (*Unfolding*)

Let $P_1$ and $P_2$ be ground programs. Program $P_2$ results from program $P_1$ by *unfolding* $(P_1 \ \mapsto_U \ P_2)$ iff there is a rule $A \leftarrow \mathcal{B}$ in $P_1$ and a positive literal $B \in \mathcal{B}$ such that

$$P_2 = \big(P_1 - \{A \leftarrow \mathcal{B}\}\big) \ \cup \ \big\{A \leftarrow \big((\mathcal{B} - \{B\}) \cup \mathcal{B}'\big) \ \big| \ B \leftarrow \mathcal{B}' \ \in \ P_1\big\}.$$

*Definition 75* (*Deletion of Non-minimal Rules*)

Let $P_1$ and $P_2$ be ground programs. Program $P_2$ results from the program $P_1$ by *deletion of non-minimal rules* $(P_1 \mapsto_M P_2)$ iff there are rules $A \leftarrow \mathcal{B}$ and $A \leftarrow \mathcal{B}'$ in $P_1$ such that $\mathcal{B} \subset \mathcal{B}'$ and $P_2 = P_1 - \{A \leftarrow \mathcal{B}'\}$.

*Lemma 76*

If a semantics $\mathcal{S}$ allows *unfolding* and the *deletion of non-minimal rules*, then it also allows *success*.

*Proof*

*Success* corresponds to *unfolding*, when the program contains a fact $B \leftarrow \emptyset$ and $B$ is replaced by $\emptyset$. However, there might be further rules about $B$ besides this fact. But then the resulting rules are certainly non-minimal, so we can remove them with $\mapsto_M$. □

*Lemma 77*

If a semantics $\mathcal{S}$ allows *unfolding*, then it also allows *failure*.

*Proof*

Obviously, the transformation $\mapsto_F$ is a very special case of *unfolding*, i.e., if there are no rules about the unfolded literal. □

*Lemma 78*

Let $\mathcal{S}$ be a semantics which allows *unfolding*, *deletion of non-minimal rules*, and *positive* and *negative reduction*. Then also

$$P_1 \mapsto_{PSNF} P_2 \implies \mathcal{S}(P_1) = \mathcal{S}(P_2)$$

for all ground programs $P_1$ and $P_2$.

*Proof*

The lemma follows immediately from the Lemmata 76 and 77. □

*Proof*

*of Theorem 15.* Strong termination is obvious, because all transformations strictly reduce the number of literals occurring in the transformed program. To show confluence, we first note that *success* commutes with the other three transformations:

If $P \ \mapsto_S \ P_1$, and $P_1 \ \mapsto_{tr} \ P'$ then $P \ \mapsto_{tr} \ P_2 \mapsto_S \ P'$

and

    If $P \mapsto_{tr} P_1$, and $P_1 \mapsto_S P'$ then $P \mapsto_S P_2 \mapsto_{tr} P'$

for $tr \in \{F, N, P\}$.

To show confluence, it suffices (using general results about confluent systems) to prove: if $P \mapsto_X P_1$ and $P \mapsto^*_{PNF} P_2$ then there is $P'$ such that $P_1 \mapsto^*_{PNF} P'$ and $P_2 \mapsto^*_{PNF} P'$, for $X \in \{F, N, P\}$. This follows immediately by checking that the following diagrams commute:

$$
\begin{array}{ccc}
P \xrightarrow{\;\mapsto^*_{PNF}\;} P_2 & \quad P \xrightarrow{\;\mapsto^*_{PNF}\;} P_2 & \quad P \xrightarrow{\;\mapsto^*_{PNF}\;} P_2 \\
{\scriptstyle P}\downarrow \qquad\qquad \downarrow{\scriptstyle P} & {\scriptstyle N}\downarrow \qquad\qquad \downarrow{\scriptstyle N} & {\scriptstyle F}\downarrow \qquad\qquad \downarrow{\scriptstyle F} \\
P_1 \xrightarrow[\;\mapsto^*_{PNF}\;]{} P' & \quad P_1 \xrightarrow[\;\mapsto^*_{PNF}\;]{} P' & \quad P_1 \xrightarrow[\;\mapsto^*_{PNF}\;]{} P'
\end{array}
$$

The diagrams commute for the following reason: the condition for the application of one of the transformations $P$, $N$, or $F$ is the existence of a fact or the non-existence of a head atom. By all our transformations no facts can be deleted and no rules with new heads can be generated. Thus, if a transformation is applicable to a rule, it will stay applicable if other transformations are applied first.

It remains to show the relation to Fitting's least fixpoint of $\Phi_P$. This operator can be best explained if we view it as a mapping between 3-valued interpretations. If we denote the set of truth values by $\mathbf{3}_k$, then $\Phi_P : \mathbf{3}_k{}^{B_P} \longmapsto \mathbf{3}_k{}^{B_P}$; $\mathcal{I} \longmapsto \Phi_P(\mathcal{I})$ is defined as follows: ($A$ is a ground atom)

$$
\Phi_P(\mathcal{I})(A) = \left\{
\begin{array}{l}
\mathbf{t}, \text{ if there is a clause } A \leftarrow L_1, \ldots, L_n \text{ in } P \text{ with:} \\
\quad\text{---}\; \forall i \leq n \text{ we have:} \\
\qquad\text{--}\; \mathcal{I}(L_i) = \mathbf{t}. \\
\mathbf{f}, \text{ if for all clauses } A \leftarrow L_1, \ldots, L_n \in P \text{ we have:} \\
\quad\text{---}\; \exists i \leq n \text{ with:} \\
\qquad\text{--}\; \mathcal{I}(L_i) = \mathbf{f}. \\
\mathbf{u}, \text{ otherwise.}
\end{array}
\right.
$$

We will identify $\mathrm{lfp}(\Phi_P)$ with a three-valued model resp. with a set of literals. Starting from the empty three-valued interpretation, where all atoms are undefined, true and false atoms are added by applying $\Phi_P$ until a fixpoint is reached.

To determine the exact relationship between the iterations of $\Phi_P$ and the applications of our transformations, we define

$$
\begin{array}{lcl}
\mathrm{step}_1(P) & := & \text{normal form of } P \text{ after applying } \mapsto^*_{SF} \text{ and } \mapsto^*_{PN} \\
\mathrm{step}_{i+1}(P) & := & \text{normal form of } \mathrm{step}_i(P) \text{ after applying } \mapsto^*_{SF} \text{ and } \mapsto^*_{PN}
\end{array}
$$

Obviously (by inspection of the definition of $\Phi_P$) we have the following result:

$$
\Phi_P \uparrow^i = \mathit{known}_{BASE(P)}(\mathrm{step}_i(P)).
$$

Clearly, $\Phi_P$ reaches a fixpoint if and only if the program $P$ is irreducible with respect to $\mapsto^*_{SFPN}$. The computation of lfp($\Phi_P$) therefore corresponds to a specific ordering of our transformations. $\quad\square$

*Proof*
*of Lemma 18.*

1. First we have to show that $\mathcal{A} = BASE(P) - \mathrm{lfp}(T_{P_1,\emptyset})$ satisfies the first requirement in Definition 17, namely that for every rule $A \leftarrow \mathcal{B}$ in $P_1$, if $A \in \mathcal{A}$, then $\mathcal{B} \cap \mathcal{A} \neq \emptyset$. Suppose that this was not the case. Then $P_1$ would contain a rule $A \leftarrow \mathcal{B}$ with $pos(\mathcal{B}) \subset \mathrm{lfp}(T_{P_1,\emptyset})$ and $A \notin \mathrm{lfp}(T_{P_1,\emptyset})$. But this is impossible, since if $pos(\mathcal{B})$ is already derived by $T_{P_1,\emptyset}$, then $A$ will be derived in the next step. Thus, *loop detection* can indeed be applied with this set $\mathcal{A}$, and we get

$$P_2 = \{ (A \leftarrow \mathcal{B}) \in P_1 \mid A \in \mathrm{lfp}(T_{P_1,\emptyset}) \}.$$

2. The second part is to show that $P_2$ is irreducible w.r.t. $\mapsto_L$. Suppose that *loop detection* is applicable to $P_2$ based on some set $\mathcal{A}'$ of ground atoms. Because of the second and third conditions in Definition 17, an atom from $\mathcal{A}'$ must appear in at least on rule head of $P_2$. By the construction of $P_2$, this atom is contained in lfp($T_{P_1,\emptyset}$). But this is impossible: We show by induction on $i$ that $\mathcal{A}' \cap (T_{P_1,\emptyset} \uparrow i) = \emptyset$. For $i = 0$ this is trivial. Let now $A \leftarrow \mathcal{B}$ be applied in the $i$-th iteration. By the induction hypothesis, we know that $pos(\mathcal{B}) \cap \mathcal{A}' = \emptyset$ and thus $\mathcal{B} \cap \mathcal{A}' = \emptyset$. Since $A \in (T_{P_1,\emptyset} \uparrow i)$, the rule is also contained in $P_2$. But then the first condition in Definition 17 implies that $A \notin \mathcal{A}'$.

$\square$

The following lemma shows that *loop detection* is a special case of *deletion of tautologies* and *unfolding*.

*Lemma 79*
If $P_1 \mapsto_L P_2$, then also $P_1 \mapsto^*_{TU} P_2$. Thus any semantics $\mathcal{S}$, which allows *unfolding* and *deletion of tautologies*, also allows *loop detection*.

*Proof*
Let $\mathcal{A}$ be a set of ground atoms such that for all rules $A \leftarrow \mathcal{B}$ in $P_1$, if $A \in \mathcal{A}$, then $\mathcal{B} \cap \mathcal{A} \neq \emptyset$. Let $P_2 := \{A \leftarrow \mathcal{B} \in P_1 \mid A \notin \mathcal{A}\}$. We will show that $P_1 - P_2$, i.e. the rules with an atom from $\mathcal{A}$ in the head, can be deleted by unfolding and elimination of tautologies, by induction on the number $n$ of atoms from $\mathcal{A}$, which actually appear in rule bodies of $P_1 - P_2$. If $n = 0$, no atom of $\mathcal{A}$ appears in a body of a rule from $P_1 - P_2$. But since each of the deleted rules must have such an atom in the body, this means $P_1 = P_2$, and thus $P_1 \mapsto^*_{TU} P_2$ trivially holds. Let now $n > 0$ and $A_0 \in \mathcal{A}$ be an atom which appears in at least one body of a rule from $P_1 - P_2$. In a first step we eliminate from $P_1$ all tautological rules containing $A_0$ in head and body. These rules are contained in $P_1 - P_2$, so we are allowed to remove them. In the second step, we unfold all remaining occurrences of $A_0$ in bodies of rules which also contain an atom from $\mathcal{A}$ in their head. Let the resulting program be called $P_1'$. Then we have the following:

- In $P_1'$, $A_0$ does not appear in any rule body with an atom from $\mathcal{A}$ in the head. Since we have eliminated the tautologies first, the rules generated by the unfolding do not contain $A_0$ in their bodies.
- The number of atoms from $A$ that appear in rules in $P_1'$ with head in $\mathcal{A}$ is at most $n - 1$.
- The rules which were deleted by the unfolding steps contain an element of $\mathcal{A}$ in the head, so they are contained in $P_1 - P_2$. The rules which were generated by the unfolding step also contain an atom from $\mathcal{A}$ in the head (since the head is not changed by unfolding). Thus, $P_2 = \{A \leftarrow \mathcal{B} \in P_1' \mid A \notin \mathcal{A}\}$.

The induction hypothesis then gives $P_1' \mapsto_{TU}^* P_2$, and thus together we have $P_1 \mapsto_{TU}^* P_1' \mapsto_{TU}^* P_2$.   □

**Theorem 80** (*Equivalence of Program Remainder*)
Let $\mathcal{S}$ be any semantics which allows *unfolding*, *deletion of tautologies*, *positive* and *negative reduction*, and the *deletion of non-minimal rules*. Then

$$\mathcal{S}(\widehat{P}) = \mathcal{S}(P)$$

for every program $P$ with remainder $\widehat{P}$.

*Proof*
This follows immediately from the Lemmata 78 and 79.   □

*Proof*
*of Theorem 23.* Since the well-founded semantics allows the transformations required to apply Theorem 80, $P$ and $\widehat{P}$ have the same well-founded model, i.e., $W_P^* = W_{\widehat{P}}^*$. Thus it suffices to show that $W_{\widehat{P}}^* = known_{BASE(P)}(\widehat{P})$. We do this by executing the alternating fixpoint procedure on $\widehat{P}$.

1. $K_0 = facts(\widehat{P})$: It is clear that at least these facts are contained in $K_0$, i.e. $facts(\widehat{P}) \subseteq K_0$. To show equality assume that $K_0 - facts(\widehat{P}) \neq \emptyset$. Let $A \leftarrow L_1 \wedge \cdots \wedge L_n$ be the first rule with non-empty body applied by the alternating fixpoint procedure. By Definition 2 the $L_i$ are all positive and contained in $facts(\widehat{P})$. But then the *success* transformation is applicable to $\widehat{P}$. This contradicts the irreducibility of $\widehat{P}$.

2. $U_0 = heads(\widehat{P})$: The inclusion $U_0 \subseteq heads(\widehat{P})$ is trivial. Now assume that $U_0 \subset heads(\widehat{P})$ holds. Let $\mathcal{A} := BASE(P) - U_0$ and let $A \leftarrow \mathcal{B}$ be an arbitrary rule of $\widehat{P}$ with $A \in \mathcal{A}$ (such a rule exists because $U_0 \neq heads(\widehat{P})$). If $\mathcal{B} \cap \mathcal{A} \neq \emptyset$, *loop detection* were applicable, contradicting the irreducibility of $\widehat{P}$. Thus we have $\mathcal{B} \cap \mathcal{A} = \emptyset$, and therefore $pos(\mathcal{B}) \subseteq U_0$. But then, since $A \in \mathcal{A}$, i.e., $A \notin U_0$, there must be a negative literal **not** $B \in \mathcal{B}$ with $B \in K_0$. In this case *negative reduction* would be applicable to $\widehat{P}$ which is impossible because $\widehat{P}$ is irreducible.

3. $K_1 = facts(\widehat{P})$: The inclusion $facts(\widehat{P}) \subseteq K_1$ is trivial, because the fixpoint $K_1 := lfp(T_{\widehat{P}, U_0})$ (cf. Definition 2) of course contains all facts of $\widehat{P}$. To show equality assume that $K_1 - facts(\widehat{P}) \neq \emptyset$. Let $A \leftarrow L_1 \wedge \cdots \wedge L_n \in \widehat{P}$ be the first rule with non-empty body applied during the computation of $K_1$. If it contains a positive body literal $L_i$, $L_i \in \mathit{facts}(\widehat{P})$ must hold. But then, *success* is applicable, contradicting the irreducibility of $\widehat{P}$. If it contains a negative body literal $L_1 = \mathbf{not}\, B$, $B \notin U_0 = \mathit{heads}(\widehat{P})$ must hold. But then, *positive reduction* is applicable, contradicting again the irreducibility of $\widehat{P}$.

4. Since $K_1 = K_0$, we also have $U_1 = U_0$, so the fixpoint is reached.

□

*Proof of Theorem 24.* Let us temporarily call the transformation on the right hand side *wred*, i.e. for any program $P'$ let

$$\mathit{wred}(P') := \big\{ A \leftarrow (\mathcal{B} - W_{P'}^*) \;\big|\; A \leftarrow \mathcal{B} \,\in\, \mathit{ground}(P')$$
$$\text{and for every } B \in \mathcal{B}\colon\, {\sim} B \notin W_{P'}^* \big\}.$$

1. We first show that

$$P_1 \mapsto_X P_2 \implies \mathit{wred}(P_1) = \mathit{wred}(P_2).$$

   (a) $P_1 \mapsto_P P_2$: *Positive reduction* removes a negative body literal $\mathbf{not}\, B$ from some rule $A \leftarrow \mathcal{B}$, if $B \notin \mathit{heads}(P_1)$. *Positive reduction* does not change the well-founded model. Furthermore $\mathbf{not}\, B$ is certainly true in the well-founded model. Thus $\mathcal{B} - W_{P_1}^* = \big(\mathcal{B} - \{\mathbf{not}\, B\}\big) - W_{P_2}^*$.

   (b) $P_1 \mapsto_N P_2$: *Negative reduction* deletes a rule $A \leftarrow \mathcal{B}$ where $\mathcal{B}$ contains a negative literal $\mathbf{not}\, B$ such that $B \in \mathit{facts}(P_1)$. But then $B$ is true in the well-founded model of $P_1$ and $P_2$, so the *wred*-transformation will delete this rule anyway.

   (c) $P_1 \mapsto_S P_2$: *Success* removes a positive body literal $B$ from some rule $A \leftarrow \mathcal{B}$, if $B \in \mathit{facts}(P_1)$. *Success* does not change the well-founded model, and furthermore $B$ is certainly true in this well-founded model. Thus we can conclude $\mathcal{B} - W_{P_1}^* = \big(\mathcal{B} - \{B\}\big) - W_{P_2}^*$.

   (d) $P_1 \mapsto_F P_2$: *Failure* deletes a rule $A \leftarrow \mathcal{B}$ where $\mathcal{B}$ contains a positive literal $B$ such that $B \notin \mathit{heads}(P_1)$. But then $\mathbf{not}\, B$ is true in the well-founded model of $P_1$ and $P_2$, so the *wred*-transformation will delete this rule anyway.

   (e) $P_1 \mapsto_L P_2$: *Loop detection* w.r.t. to an unfounded set $\mathcal{A}$ deletes all rules $A \leftarrow \mathcal{B}$ where $A \in \mathcal{A}$. By the first condition of Definition 17, this also means that $\mathcal{B}$ contains a positive body literal $B$ with $B \in \mathcal{A}$. But since all atoms in the unfounded set $\mathcal{A}$ are false in the well-founded model of $P_1$ and $P_2$, the *wred*-transformation will delete these rules anyway.

2. Since $\mathit{ground}(P) \mapsto_X^* \widehat{P}$, we have $\mathit{wred}(\mathit{ground}(P)) = \mathit{wred}(\widehat{P})$. Because the well-founded semantics is instantiation-invariant the definition of *wred* implies $\mathit{wred}(P) = \mathit{wred}(\widehat{P})$. We now only have to show $\mathit{wred}(\widehat{P}) = \widehat{P}$. Assume that $\mathit{wred}(\widehat{P}) \neq \widehat{P}$. Then there is a rule $A \leftarrow \mathcal{B} \in \widehat{P}$ such that there is $B \in \mathcal{B}$ and either $A \leftarrow \mathcal{B}$ has been modified by *wred* because $B \in W_{\widehat{P}}^*$ or entirely deleted because ${\sim} B \in W_{\widehat{P}}^*$. But we know by Theorem 23 that

$$W_{\widehat{P}}^* = \mathit{known}_{BASE(P)}(\widehat{P}).$$

We can distinguish the following four cases:

(a) $B \in W_{\widehat{P}}^*$.

    i  $B$ is positive. It follows that $B \in facts(\widehat{P})$. But then *success* is applicable, contradicting the irreducibility of $\widehat{P}$.

    ii  $B = \mathbf{not}\, C$ is negative. It follows that $C \notin heads(\widehat{P})$. But then *positive reduction* is applicable, again contradicting the irreducibility of $\widehat{P}$.

(b) $\sim B \in W_{\widehat{P}}^*$.

    i  $B$ is positive. It follows that $\mathbf{not}\, B \in W_{\widehat{P}}^*$ and thus $B \notin heads(\widehat{P})$. But then *failure* is applicable, contradicting the irreducibility of $\widehat{P}$.

    ii  $B = \mathbf{not}\, C$ is negative. It follows that $C \in W_{\widehat{P}}^*$ and thus $C \in facts(\widehat{P})$. But then *negative reduction* is applicable, again contradicting the irreducibility of $\widehat{P}$.

□

## B Proofs of Section 5

*Proof*
*of Lemma 31.* Following Lemma 18, $\mathcal{A} := BASE(P) - \mathrm{lfp}(T_{P,\emptyset})$ is a valid reference set for *loop detection*. Let $P'$ be the result of applying *loop detection* w.r.t. this set $\mathcal{A}$ to $ground(P)$. It follows that

$$P' = \{A \leftarrow \mathcal{B} \in ground(P) \mid A \in \mathrm{lfp}(T_{P,\emptyset})\}.$$

Thus $heads(P') \subseteq \mathrm{lfp}(T_{P,\emptyset})$, and we can use *failure* to remove also the rules which contain a positive body literal $B$ with $B \notin \mathrm{lfp}(T_{P,\emptyset})$. Then the result is

$$P'' = \{A \leftarrow \mathcal{B} \in ground(P) \mid \{A\} \cup pos(\mathcal{B}) \subseteq \mathrm{lfp}(T_{P,\emptyset})\}.$$

As mentioned above, it is obvious that $\mathrm{lfp}(T_{P,\emptyset}) = heads(\mathrm{lfp}(\bar{T}_P))$. To prove $P'' = \mathrm{lfp}(\bar{T}_P)$, we show that $P'' \subseteq \mathrm{lfp}(\bar{T}_P)$ and $P'' \supseteq \mathrm{lfp}(\bar{T}_P)$:

1. It is clear that $A \leftarrow \mathcal{B} \in ground(P)$ with

$$pos(\mathcal{B}) \subseteq \mathrm{lfp}(T_{P,\emptyset}) = heads(\mathrm{lfp}(\bar{T}_P)),$$

    thus $A \leftarrow \mathcal{B}$ must also be contained in $\mathrm{lfp}(\bar{T}_P)$.

2. If $A \leftarrow \mathcal{B} \in ground(P)$ is derived by $\bar{T}_P$, this means that

$$pos(\mathcal{B}) \subseteq heads(\mathrm{lfp}(\bar{T}_P)) = \mathrm{lfp}(T_{P,\emptyset})$$

    and $A \in heads(\mathrm{lfp}(\bar{T}_P)) = \mathrm{lfp}(T_{P,\emptyset})$.

□

## C Proofs of Section 6

*Proof*

*of Lemma 40.* To begin with, we recall Lemma 31 and observe that $\bar{K}_0$ is well-defined. The proof by induction is based on the following reasoning:

1. As one possible application of $\mapsto_{PS}$ never interferes with another possible application of $\mapsto_{PS}$, the transformations $\mapsto_P$ and $\mapsto_S$ can be applied in any order and always produce the same result. In particular, it is legal to apply all possible $\mapsto_P$ transformations first, followed by all possible $\mapsto_S$ transformations, because $\mapsto_S$ does not derive any new negative information, i.e., it does not delete any rule instance.

2. Analogously, to construct the normal form w.r.t. $\mapsto_{NLF}$ it is legal to apply all possible $\mapsto_N$ transformations first before applying $\mapsto_{LF}$, since *loop detection* and *failure* do not derive any new facts, i.e., they only delete rule instances. All applicable $\mapsto_N$ transformations can be applied in any order because they do not interfere with each other and together always delete the same set of rule instances. Since the greatest unfounded set of a program is the union of all unfounded sets, the normal form w.r.t. $\mapsto_L$ can be constructed by one application of $\mapsto_L$ w.r.t. this greatest unfounded set (cf. Lemma 18). Finally, rule instances with positive body literals known to be false are deleted by applying $\mapsto_F$. If for an atom $A$ all rule instances having $A$ as their head atom can be deleted by $\mapsto_F$ then $\{A\}$ is an unfounded set and the corresponding rules have already been deleted by $\mapsto_L$. Thus all possible applications of $\mapsto_F$ can be performed in any order.

□

*Lemma 81*
Let $P_1$ be any ground program and let $P_2$ be the normal form of $P_1$ w.r.t. $\mapsto_S$ (*success*). Then:

$$P_2 = \big\{ A \leftarrow (\mathcal{B} - facts(P_2)) \wedge \mathbf{not}\,\mathcal{C} \ \big|\ A \leftarrow \mathcal{B} \wedge \mathbf{not}\,\mathcal{C} \ \in\ P_1 \big\}.$$

*Proof*
Every rule in $P_2$ results from a rule of $P_1$ by removing positive body literals. Only positive body literals can be removed which are known as facts, and the set of facts can only increase during this process. Since $P_2$ is irreducible w.r.t. $\mapsto_S$, atoms in $facts(P_2)$ cannot occur any longer as positive body literals. □

*Lemma 82*
Let $P_1$ be any ground program and let $P_2$ be the normal form of $P_1$ w.r.t. $\mapsto_N$ (*negative reduction*). Then:

$$P_2 = \big\{ A \leftarrow \mathcal{B} \wedge \mathbf{not}\,\mathcal{C} \ \in\ P_1 \ \big|\ \mathcal{C} \cap facts(P_1) = \emptyset \big\}.$$

*Proof*
*Negative reduction* deletes the rules which have a negative body literal conflicting with a fact. The set of facts is not changed during this process, i.e. $facts(P_1) = facts(P_2)$. □

*Lemma 83*

Let $P_1$ be any ground program and let $P_2$ be the normal form of $P_1$ w.r.t. $\mapsto_P$ (*positive reduction*). Then:

$$P_2 = \big\{A \leftarrow \mathcal{B} \wedge \mathbf{not}(\mathcal{C} \cap \textit{heads}(P_1)) \; \big| \; A \leftarrow \mathcal{B} \wedge \mathbf{not}\,\mathcal{C} \; \in \; P_1\big\}.$$

*Proof*

*Positive reduction* removes the negative body literals for which the corresponding positive atom does not appear as a head literal. The set of head literals is not changed during this process, i.e. $\textit{heads}(P_1) = \textit{heads}(P_2)$.   $\square$

*Lemma 84*

Let $P_1$ be any ground program and let $P_2$ be the normal form of $P_1$ w.r.t. $\mapsto_{LF}$ (*loop detection* and *failure*). Then: $P_2 = \mathrm{lfp}(\bar{T}_{P_1})$.

*Proof*

1.  $P_1 \mapsto^*_{LF} \mathrm{lfp}(\bar{T}_{P_1})$ has already been proven in Lemma 31.
2.  We show that $\mathrm{lfp}(\bar{T}_{P_1})$ is irreducible w.r.t. $\mapsto_{LF}$. Let us assume that $\mapsto_L$ is applicable. Let $\mathcal{A}$ be as in the definition of $\mapsto_L$ (Definition 17). Consider the first iteration of the fixpoint iteration in which a rule instance $A \leftarrow \mathcal{B} \wedge \mathbf{not}\,\mathcal{C}$ is derived with $\mathcal{B} \cap \mathcal{A} \neq \emptyset$. There must be such a rule instance since *loop detection* would not be applicable otherwise. Let $B \in \mathcal{B} \cap \mathcal{A}$. Then a rule instance $B \leftarrow \mathcal{B}' \wedge \mathbf{not}\,\mathcal{C}$ has been derived in a previous iteration. This means that $\mathcal{B}' \cap \mathcal{A} = \emptyset$. But this contradicts the first condition in Definition 17 which implies that $\mapsto_L$ is not applicable to $\mathrm{lfp}(\bar{T}_{P_1})$. Now let us assume that $\mapsto_F$ is applicable. Then there is a rule instance $A \leftarrow \mathcal{B} \wedge \mathbf{not}\,\mathcal{C}$ and an atom $B \in \mathcal{B}$ that is not contained in $\textit{heads}(\mathrm{lfp}(\bar{T}_{P_1}))$. But this contradicts the iterative construction of $\mathrm{lfp}(\bar{T}_{P_1})$.

    $\square$

*Lemma 85*

Let $P_1$ be any ground program and let $P_2$ be the normal form of $P_1$ w.r.t. $\mapsto_{LF}$ (*loop detection* and *failure*). Then:

$$P_2 = \big\{A \leftarrow \mathcal{B} \wedge \mathbf{not}\,\mathcal{C} \; \in \; P_1 \; \big| \; \mathcal{B} \subseteq \textit{heads}(P_2)\big\}.$$

*Proof*

This follows immediately from Lemma 84: Since $P_2$ is a fixpoint of $\bar{T}_{P_1}$, every rule $A \leftarrow \mathcal{B} \wedge \mathbf{not}\,\mathcal{C} \; \in \; P_1$ with $\mathcal{B} \subseteq \textit{heads}(P_2)$ must be contained in $P_2$. On the other hand, it follows from the iterative construction of $\mathrm{lfp}(\bar{T}_{P_1})$ that any rule $A \leftarrow \mathcal{B} \wedge \mathbf{not}\,\mathcal{C}$ contained in $\mathrm{lfp}(\bar{T}_{P_1})$ must satisfy $\mathcal{B} \subseteq \textit{heads}(\mathrm{lfp}(\bar{T}_{P_1}))$.   $\square$

*Proof*

*of Theorem 42.* We do not only show $\textit{facts}(\bar{K}_i) = K_i$ and $\textit{heads}(\bar{U}_i) = U_i$ but also the following stronger characterizations of $\bar{K}_i$ and $\bar{U}_i$:

$$
\begin{aligned}
\bar{K}_0 \;=\; & \big\{A \leftarrow (\mathcal{B} - K_0) \wedge \mathbf{not}\,\mathcal{C} \; \big| \\
& \quad A \leftarrow \mathcal{B} \wedge \mathbf{not}\,\mathcal{C} \in \textit{ground}(P),\ \mathcal{B} \subseteq \mathrm{lfp}(T_{P,\emptyset})\big\} \\
\bar{U}_0 \;=\; & \big\{A \leftarrow (\mathcal{B} - K_0) \wedge \mathbf{not}\,\mathcal{C} \; \big| \\
& \quad A \leftarrow \mathcal{B} \wedge \mathbf{not}\,\mathcal{C} \in \textit{ground}(P),\ \mathcal{B} \subseteq U_0,\ \mathcal{C} \cap K_0 = \emptyset\big\}
\end{aligned}
$$

$$\begin{aligned}
\bar{K}_i \;=\; & \big\{ A \leftarrow (\mathcal{B} - K_i) \wedge \mathbf{not}(\mathcal{C} \cap U_{i-1}) \;\big| \\
& \qquad A \leftarrow \mathcal{B} \wedge \mathbf{not}\,\mathcal{C} \in ground(P),\; \mathcal{B} \subseteq U_{i-1},\; \mathcal{C} \cap K_{i-1} = \emptyset \big\},\; i > 0 \\
\bar{U}_i \;=\; & \big\{ A \leftarrow (\mathcal{B} - K_i) \wedge \mathbf{not}(\mathcal{C} \cap U_{i-1}) \;\big| \\
& \qquad A \leftarrow \mathcal{B} \wedge \mathbf{not}\,\mathcal{C} \in ground(P),\; \mathcal{B} \subseteq U_i,\; \mathcal{C} \cap K_i = \emptyset \big\},\; i > 0.
\end{aligned}$$

The proof is by induction on $i$.

1. **Proof for $\bar{K}_0$:** Due to Lemma 31 we already know that $ground(P) \mapsto_{LF} \mathrm{lfp}(\bar{T}_P)$. It is easy to show by induction on the number of derivation steps that

$$\mathrm{lfp}(\bar{T}_P) = \big\{ A \leftarrow \mathcal{B} \wedge \mathbf{not}\,\mathcal{C} \;\big|\; A \leftarrow \mathcal{B} \wedge \mathbf{not}\,\mathcal{C} \in ground(P),\; \mathcal{B} \subseteq T_{P,\emptyset} \big\}$$

   (because $heads(\bar{T}_P \uparrow j) = T_{P,\emptyset} \uparrow j$). By Definition 39 $\bar{K}_0$ results from $\mathrm{lfp}(\bar{T}_P)$ by applying $\mapsto_S$ until irreducibility w.r.t. $\mapsto_S$. We show that $facts(\bar{K}_0) = K_0$:

   (a) $K_0 \subseteq facts(\bar{K}_0)$: We show $T_{P^+} \uparrow j \subseteq facts(\bar{K}_0)$ by induction on $j$: The case $j = 0$ is trivial since $T_{P^+} \uparrow 0 = \emptyset$. Now suppose that $A$ has been derived in step $j$ by applying $A \leftarrow \mathcal{B} \in ground(P^+)$. Then $\mathcal{B} \subseteq T_{P^+} \uparrow (j-1)$ and therefore $\mathcal{B} \subseteq facts(\bar{K}_0)$ by the induction hypothesis. Furthermore we know that $T_{P^+} \uparrow (j-1) \subseteq \mathrm{lfp}(T_{P^+}) \subseteq \mathrm{lfp}(T_{P,\emptyset})$ and therefore $A \leftarrow \mathcal{B} \in \mathrm{lfp}(\bar{T}_P)$. But since $\mathcal{B} \subseteq facts(\bar{K}_0)$ and $\bar{K}_0$ is the normal form of $\mathrm{lfp}(\bar{T}_P)$ w.r.t. $\mapsto_S$, it follows that $A \in facts(\bar{K}_0)$.

   (b) $facts(\bar{K}_0) \subseteq K_0$: The proof is by induction on the number of applications of $\mapsto_S$. All facts contained in $\mathrm{lfp}(\bar{T}_P)$ are also contained in $ground(P^+)$ and thus in $K_0$. Suppose that $A$ is derived from the rule $A \leftarrow \mathcal{B} \in \mathrm{lfp}(\bar{T}_P)$ by $j$ applications of $\mapsto_S$. Then the atoms in $\mathcal{B}$ are derived in fewer steps, and thus $\mathcal{B} \subseteq K_0$ by the induction hypothesis. Since $K_0$ is a fixpoint of $T_{P^+}$ and $A \leftarrow \mathcal{B} \in ground(P^+)$ it follows that $A \in K_0$.

   The characterization of $\bar{K}_0$ now follows from $facts(\bar{K}_0) = K_0$ by Lemma 81 and the introductory remark on $\mathrm{lfp}(\bar{T}_P)$.

2. **Proof for $\bar{U}_0$:** To construct $\bar{U}_0$ from $\bar{K}_0$ we may compute the normal form of $\bar{K}_0$ w.r.t. $\mapsto_N$ first (see proof of Lemma 40) which we denote by $\bar{N}_0$ in the following. By Lemma 82 and the characterization of $\bar{K}_0$ already shown above

$$\bar{N}_0 \;=\; \big\{ A \leftarrow (\mathcal{B} - K_0) \wedge \mathbf{not}\,\mathcal{C} \;\big|\; A \leftarrow \mathcal{B} \wedge \mathbf{not}\,\mathcal{C} \in ground(P),\; \mathcal{B} \subseteq \mathrm{lfp}(T_{P,\emptyset}),\; \mathcal{C} \cap K_0 = \emptyset \big\}$$

   holds. Let $\bar{U}_0$ be the normal form of $\bar{N}_0$ w.r.t. $\mapsto_{LF}$. We show that $heads(\bar{U}_0) = U_0$:

   (a) $U_0 \subseteq heads(\bar{U}_0)$: Because we have $U_0 = \mathrm{lfp}(T_{P,K_0})$, we prove $T_{P,K_0} \uparrow j \subseteq heads(\bar{U}_0)$ by induction on $j$. The case $j = 0$ is trivial. Let now $A \in T_{P,K_0} \uparrow j$ have been derived by using the rule instance $A \leftarrow \mathcal{B} \wedge \mathbf{not}\,\mathcal{C}$. Then $\mathcal{B} \subseteq T_{P,K_0} \uparrow (j-1)$ and $\mathcal{C} \cap K_0 = \emptyset$. Also $T_{P,K_0} \uparrow (j-1) \subseteq \mathrm{lfp}(T_{P,K_0}) \subseteq \mathrm{lfp}(T_{P,\emptyset})$, thus $A \leftarrow (\mathcal{B} - K_0) \wedge \mathbf{not}\,\mathcal{C}$ is contained in $\bar{N}_0$. By the induction hypothesis, we have $\mathcal{B} \subseteq heads(\bar{U}_0)$. By Lemma 85 this implies that $A \leftarrow (\mathcal{B} - K_0) \wedge \mathbf{not}\,\mathcal{C}$ is not deleted by *loop detection* or *failure*. Thus $A \in heads(\bar{U}_0)$.

   (b) $heads(\bar{U}_0) \subseteq U_0$: By Lemma 84, $\bar{U}_0 = \mathrm{lfp}(\bar{T}_{\bar{N}_0})$. Therefore, we show

$heads(\bar{T}_{\bar{N}_0} \uparrow j) \subseteq U_0$ by induction on $j$. The case $j = 0$ is trivial again. Now suppose that $A \leftarrow (\mathcal{B} - K_0) \land \mathbf{not}\,\mathcal{C}$ has been derived in step $j$. Then $(\mathcal{B} - K_0) \subseteq heads(\bar{T}_{\bar{N}_0} \uparrow (j-1))$ and therefore $(\mathcal{B} - K_0) \subseteq U_0$ by the induction hypothesis. But $K_0 \subseteq U_0$ also holds by Lemma 4, thus $\mathcal{B} \subseteq U_0$. We also have $\mathcal{C} \cap K_0 = \emptyset$, otherwise the rule instance would not be contained in $\bar{N}_0$. But since $U_0 = \mathrm{lfp}(T_{P,K_0})$, it follows that $A \in U_0$.

Our characterization of $\bar{U}_0$ follows from $heads(\bar{U}_0) = U_0$ by Lemma 85.

3. **Proof for** $\bar{K}_i$, $i \geq 1$**:** To construct $\bar{K}_i$, we first apply *positive reduction* to $\bar{U}_{i-1}$ until irreducibility w.r.t. $\mapsto_P$ and denote the resulting set by $\bar{N}_i$. By the induction hypothesis we have $heads(\bar{U}_{i-1}) = U_{i-1}$ and

$$\bar{U}_{i-1} = \; \big\{ A \leftarrow (\mathcal{B} - K_{i-1}) \land \mathbf{not}(\mathcal{C} \cap U_{i-2}) \; \big|$$
$$A \leftarrow \mathcal{B} \land \mathbf{not}\,\mathcal{C} \in ground(P),\ \mathcal{B} \subseteq U_{i-1},\ \mathcal{C} \cap K_{i-1} = \emptyset \big\}.$$

By Lemma 83 and since $U_{i-1} \subseteq U_{i-2}$ (Lemma 4) it follows that

$$\bar{N}_i = \; \big\{ A \leftarrow (\mathcal{B} - K_{i-1}) \land \mathbf{not}(\mathcal{C} \cap U_{i-1}) \; \big|$$
$$A \leftarrow \mathcal{B} \land \mathbf{not}\,\mathcal{C} \in ground(P),\ \mathcal{B} \subseteq U_{i-1},\ \mathcal{C} \cap K_{i-1} = \emptyset \big\}$$

holds. Now we apply $\mapsto_S$ until irreducibility w.r.t. $\mapsto_S$, the result is $\bar{K}_i$. We show that $facts(\bar{K}_i) = K_i$:

(a) $facts(\bar{K}_i) \subseteq K_i$: The proof is by induction on the number of applications of $\mapsto_S$. Let $A \in facts(\bar{K}_i)$. Either $A \in facts(\bar{K}_{i-1})$ and the main induction hypothesis gives $A \in K_{i-1} \subseteq K_i$ or $A$ results from a rule

$$A \leftarrow (\mathcal{B} - K_{i-1}) \land \mathbf{not}(\mathcal{C} \cap U_{i-1})$$

in $\bar{N}_i$ where $\mathcal{C} \cap U_{i-1} = \emptyset$ and $\mathcal{B} - K_{i-1}$ can be derived as facts with fewer applications of $\mapsto_S$. Then the induction hypothesis gives $(\mathcal{B} - K_{i-1}) \subseteq K_i$. Furthermore $K_{i-1} \subseteq K_i$ by Lemma 4. Thus $\mathcal{B} \subseteq K_i$. But then the rule instance $A \leftarrow \mathcal{B} \land \mathbf{not}\,\mathcal{C}$ is applicable in the computation of $\mathrm{lfp}(T_{P,U_{i-1}})$ and we get $A \in K_i$.

(b) $K_i \subseteq facts(\bar{K}_i)$: Because of the equality $K_i = \mathrm{lfp}(T_{P,U_{i-1}})$, we show $(T_{P,U_{i-1}} \uparrow j) \subseteq facts(\bar{K}_i)$ by induction on $j$: The case $j = 0$ is trivial. Now suppose that $A$ has been derived in step $j$ by applying $A \leftarrow \mathcal{B} \land \mathbf{not}\,\mathcal{C}$. Then $\mathcal{B} \subseteq (T_{P,U_{i-1}} \uparrow (j-1))$ and $\mathcal{C} \cap U_{i-1} = \emptyset$. The first part gives us $\mathcal{B} \subseteq K_i$ which together with Lemma 4 implies $\mathcal{B} \subseteq K_i \subseteq U_i \subseteq U_{i-1}$. $\mathcal{C} \cap U_{i-1} = \emptyset$ implies, again using Lemma 4, that $\mathcal{C} \cap K_{i-1} = \emptyset$. Together we can conclude that

$$A \leftarrow (\mathcal{B} - K_{i-1}) \land \mathbf{not}(\mathcal{C} \cap U_{i-1})$$

is contained in $\bar{N}_i$. Moreover, $\mathcal{C} \cap U_{i-1} = \emptyset$ shows that there are actually no negative body literals, so the rule applied is $A \leftarrow (\mathcal{B} - K_{i-1})$. $\mathcal{B} \subseteq (T_{P,U_{i-1}} \uparrow (j-1))$ together with the induction hypothesis gives us $\mathcal{B} \subseteq facts(\bar{K}_i)$. But since $\bar{K}_i$ is irreducible w.r.t. $\mapsto_S$, and $A \leftarrow (\mathcal{B} - K_{i-1})$ is contained in $\bar{N}_i$, we can finally conclude $A \in \bar{K}_i$.

Our characterization now follows from $facts(\bar{K}_i) = K_i$ by Lemma 81 and Lemma 4 ($K_{i-1} \subseteq K_i$).

4. **Proof for** $\bar{U}_i$, $i \geq 1$: To construct $\bar{U}_i$ from $\bar{K}_i$, we first compute the normal form of $\bar{K}_i$ w.r.t. $\mapsto_N$ which we denote by $\bar{N}_i$. By Lemma 82

$$\bar{N}_i = \big\{ A \leftarrow \mathcal{B} \wedge \mathbf{not}\, \mathcal{C} \in \bar{K}_i \ \big|\ \mathcal{C} \cap facts(\bar{K}_i) = \emptyset \big\}.$$

As we have just seen, the induction hypothesis implies the characterization of $\bar{K}_i$ and $facts(\bar{K}_i) = K_i$. Therefore and since $K_{i-1} \subseteq K_i$ (see Lemma 4) we can conclude that

$$\bar{N}_i = \big\{ A \leftarrow (\mathcal{B} - K_i) \wedge \mathbf{not}(\mathcal{C} \cap U_{i-1}) \ \big|$$
$$A \leftarrow \mathcal{B} \wedge \mathbf{not}\, \mathcal{C} \in ground(P),\ \mathcal{B} \subseteq U_{i-1},\ \mathcal{C} \cap K_i = \emptyset \big\}$$

holds. Let $\bar{U}_i$ be the normal form of $\bar{N}_i$ w.r.t. $\mapsto_{LF}$. We show $heads(\bar{U}_i) = U_i$:

(a) $heads(\bar{U}_i) \subseteq U_i$: By Lemma 84, $\bar{U}_i = \mathrm{lfp}(\bar{T}_{\bar{N}_i})$. Therefore, we show $heads(\bar{T}_{\bar{N}_i} \uparrow j) \subseteq U_i$ by induction on $j$. The case $j = 0$ is trivial. Now suppose that

$$A \leftarrow (\mathcal{B} - K_i) \wedge \mathbf{not}(\mathcal{C} \cap U_{i-1})$$

has been derived in step $j$. Then $(\mathcal{B} - K_i) \subseteq heads(\bar{T}_{\bar{N}_i} \uparrow (j-1))$ and therefore $(\mathcal{B} - K_i) \subseteq U_i$ by the induction hypothesis. But $K_i \subseteq U_i$ also holds by Lemma 4, and thus $\mathcal{B} \subseteq U_i$. We also have $\mathcal{C} \cap K_i = \emptyset$, otherwise the derived ground rule would not be contained in $\bar{N}_i$. But $U_i = \mathrm{lfp}(T_{P,K_i})$, and we can apply $A \leftarrow \mathcal{B} \wedge \mathbf{not}\, \mathcal{C}$ in the computation of $\mathrm{lfp}(T_{P,K_i})$, so $A \in U_i$.

(b) $U_i \subseteq heads(\bar{U}_i)$: Since $U_i = \mathrm{lfp}(T_{P,K_i})$, we prove

$$(T_{P,K_i} \uparrow j) \subseteq heads(\bar{U}_i)$$

by induction on $j$. The case $j = 0$ is trivial. Let now $A \in (T_{P,K_i} \uparrow j)$ be derived by applying the rule instance $A \leftarrow \mathcal{B} \wedge \mathbf{not}\, \mathcal{C}$. Then $\mathcal{B} \subseteq (T_{P,K_i} \uparrow (j-1))$ and $\mathcal{C} \cap K_i = \emptyset$. Since $T_{P,K_i} \uparrow (j-1) \subseteq \mathrm{lfp}(T_{P,K_i}) = U_i$ and $U_i \subseteq U_{i-1}$ by Lemma 4, we get $\mathcal{B} \subseteq U_{i-1}$. Therefore, the ground rule

$$A \leftarrow (\mathcal{B} - K_i) \wedge \mathbf{not}(\mathcal{C} \cap U_{i-1})$$

is contained in $\bar{N}_i$. From $\mathcal{B} \subseteq (T_{P,K_i} \uparrow (j-1))$ the induction hypothesis gives us $\mathcal{B} \subseteq heads(\bar{U}_i)$. But then $A \in heads(\bar{U}_i)$ follows from Lemma 85.

Our characterization of $\bar{U}_i$ follows from $heads(\bar{U}_i) = U_i$ by Lemma 85 and Lemma 4 ($U_i \subseteq U_{i-1}$).

$\square$